\documentclass[groupedaddress,superscriptaddress,aps,prd,preprintnumbers,onecolumn,11pt,fleqn,nofootinbib,nobibnote,eqsecnum,floatfix,showpacs]{revtex4-1}

\usepackage{graphicx}
\usepackage{bm,amsmath,amssymb}
\usepackage[mathscr]{eucal}

\long\def\comment#1{ }
\newcommand{\eqn}[1]{Eq.~\eqref{#1}}
\newcommand{\beq}{\begin{equation}}
\newcommand{\eeq}{\end{equation}}
\newcommand{\nn}{\nonumber\\}
\newcommand{\dif}{{\rm d}}

\newcommand{\rme}{{\rm e}}
\newcommand{\rmi}{{\rm i}}

\newcommand{\rmR}{{\rm R}}

\newcommand{\del}{\partial}

\newcommand{\mcal}{\mathcal}

\newcommand{\bk}{\bm{k}}

\newcommand{\bp}{\bm{p}}
\newcommand{\bx}{\bm{x}}
\newcommand{\by}{\bm{y}}

\begin{document}

\preprint{arXiv:1403.1184}

\title{The Boltzmann Equation in Classical Yang-Mills Theory}

\author{V.~Mathieu}
\email{mathieuv@indiana.edu}
\affiliation{Department of Physics, Indiana University, Bloomington, IN 47405, USA}
\affiliation{Center for Exploration of Energy and Matter, Indiana University, Bloomington, IN 47403, USA}

\author{A.H.~Mueller}
\email{amh@phys.columbia.edu}
\affiliation{Department of Physics, Columbia University, New York, NY 10027, USA}

\author{D.N.~Triantafyllopoulos}
\email{trianta@ectstar.eu}
\affiliation{European Centre for Theoretical Studies in Nuclear Physics and Related Areas (ECT*)\\
and Fondazione Bruno Kessler, Strada delle Tabarelle 286, I-38123 Villazzano (TN), Italy}

\date{\today}

\begin{abstract}
We give a detailed derivation of the Boltzmann equation, and in particular its collision integral, in classical field theory. We first carry this out in a scalar theory with both cubic and quartic interactions and subsequently in a Yang-Mills theory. Our method does not rely on a doubling of the fields, rather it is based on a diagrammatic approach representing the classical solution to the problem.  
\end{abstract}

%\keywords{}

\maketitle

%\tableofcontents

\section{\label{sec:intro}Introduction, motivation and the Boltzmann equation}

Transport phenomena in QCD matter have been the subject of extensive research over the last three decades. Particular attention has been paid to calculating quantities like conductivity, viscosity and baryon diffusion \cite{Baym:1990uj,Jeon:1995zm,Arnold:2000dr,Arnold:2003zc} or the relaxation of colorful excitations \cite{Selikhov:1993ns,Heiselberg:1994px,Bodeker:1998hm,Arnold:1998cy,Blaizot:1999xk,Blaizot:2001nr} in a weakly coupled Quark-Gluon Plasma (QGP). A key element in such studies has been the use of kinetic equations which are of the Boltzmann type. The Boltzmann equation is an equation which describes the time-evolution of occupation numbers. An occupation number is a dimensionless quantity defined as the number of particles of a given species per unit phase space and divided by the number of choices for each possible discrete degree of freedom. For example, in a $SU(N_c)$ pure gauge theory one divides by $2 (N_c^2-1)$ for the polarizations and colors of the gauge bosons to which we shall refer as gluons. The Boltzmann equation for the gluon occupation number $f(\bp,\bx,t)$ reads
 \beq
 \label{beintro}
 \bigg(
 \frac{\del}{\del t} +
 \bm{v}_{\bp} \cdot 
 \frac{\del}{\del \bx}
 +\bm{F}_{\rm ext}\cdot
 \frac{\del}{\del \bp}
 \bigg)
 f(\bp,\bx,t) = C[f],
 \eeq
with $\bm{v}_{\bp} = \bp/E_{\bp}$ the gluon velocity having unit magnitude, $\bm{F}_{\rm ext}$ a generic external force and $C[f]$ the \emph{collision term} or collision integral accounting for the interactions among gluons. Considering only $2\to 2$ elastic scattering this collision term reads
 \begin{align}
 \label{cf}
 C[f] =
 \frac{1}{4 E_{\bp}} 
 \int & 
 \widetilde{\dif \bp_1}\,
 \widetilde{\dif \bp_2}\,
 \widetilde{\dif \bp_3}\,
 (2\pi)^4 \delta^{(4)}(\Delta p)\,
 \frac{|\mcal{M}|^2_{\rm YM}}{2(N_c^2-1)}
 \nn  
 &\times\big[ 
 f_{\bp_2}f_{\bp_3}
 \big(1+f_{\bp_1}\big)
 \big(1 +f_{\bp}\big)   
 -f_{\bp}f_{\bp_1}
 \big(1+f_{\bp_2}\big)
 \big(1+f_{\bp_3}\big) 
 \big],
 \end{align}
where we have used the compact notation $f_{\bp} = f(\bp,\bx,t)$ since the integrand is local in both $\bx$ and $t$ and defined in general the integration measure
 \beq
 \label{dptilde}
 \widetilde{\dif \bp} \equiv
 \frac{\dif^3 \bp}{(2\pi)^3 2 E_{\bp}}.
 \eeq  
Energy-momentum conservation in \eqn{beintro} is explicit, while the scattering amplitude squared $|\mcal{M}|^2_{\rm YM}$ for the process $p_2 p_3 \to p p_1 $ is summed over initial and final colors and polarizations and is given below in \eqn{m2ym}. Each of the two terms in the square bracket in \eqn{cf} has an intuitive interpretation. The first is a gain term proportional to $f_{\bp_2}f_{\bp_3}$, with $p_2$ and $p_3$ disappearing to create $p$ and $p_1$, while $\big(1+f_{\bp_1}\big)$ and $\big(1 +f_{\bp}\big)$ are Bose enhancement factors. Similarly, the second is a loss term describing the disappearance of $p$ and $p_1$ in order to create $p_2$ and $p_3$. Notice also that this square bracket vanishes when occupation numbers are given by the Bose-Einstein distribution. Further aspects of this collision integral will be discussed in the next sections.

A valid question that one immediately asks is how such a kinetic equation can be derived from first principles, i.e.~from the underlying quantum field theory. Indeed, this was first addressed long time ago in non-relativistic quantum field theory \cite{Kadanoff:1962aaaa}. Using the Schwinger-Keldysh formalism and writing Dyson-Schwinger equations for the propagators, an appropriate truncation supplemented with a gradient expansion led to the non-relativistic version of the Boltzmann equation given above. Notice that in such a limit the Bose enhancement factors in the collision integral are absent and the collision integral vanishes when occupation numbers are given by the Maxwell-Boltzmann distribution. Using similar Green's function techniques in relativistic quantum field theories, the Boltzmann equation was derived in \cite{Calzetta:1986cq} for scalar fields, in \cite{Mrowczynski:1989bu} for charged scalar fields and in \cite{Mrowczynski:1992hq} for nuclear matter described by the Walecka model. A somewhat different derivation based on resuming ladder diagrams, again in a scalar field theory, was given in \cite{Jeon:1994if}, while kinetic equations for colorful excitations in a weakly coupled QGP were obtained in \cite{Blaizot:1993be,Iancu:1998sg,Blaizot:1999xk} by performing gauge covariant gradient expansions. For both a pedagogical introduction and an overview we refer the reader to \cite{Blaizot:2001nr,Calzetta:2008aaaa}.

Typically, the essential assumptions for arriving at such a kinetic equation are two. First one needs that occupation numbers do not become very large; for example in QCD one needs $f_{\bp} \ll 1/\alpha_s$ while in a scalar theory with quartic interactions ($\lambda \phi^4$ theory) this constraint would be $f_{\bp} \ll 1/\lambda$. This is necessary, since otherwise a description using on-shell scattering of individual particles no longer makes sense as the time between scatterings is too short for an on-shell approximation to be valid. Second one has to assume that there are no large wavelength modes comparable to the mean free path, otherwise one has to treat them in a suitable way. 

Here we would like to study the conditions under which bulk matter can be described by a Boltzmann equation with a collision term given by elastic scattering, but also under the additional assumption that the physical system is classical\footnote{A different connection between the classical approximation to statistical field theory and the transport theory appears in studies of baryon number violation via topological transitions in hot QCD; in that context, the quantum Boltzmann equation for the relaxation of colorful excitations has been used to construct a classical effective theory for the ``ultrasoft'' modes responsible for the topological transititons \cite{Bodeker:1998hm,Iancu:1998sg,Litim:1999ns,Litim:2001db}.}. Then the extra condition $f_{\bp} \gg 1$ is required in order to have the possibility of a quantum-classical correspondence, but when the coupling is sufficiently small there is a parametrically large window in which a kinetic description via a Boltzmann equation should be valid. In fact such an observation and the corresponding derivation have been already done a few years ago in the context of a $\lambda \phi^4$ theory \cite{Mueller:2002gd} (see also \cite{Jeon:2004dh}). 
In that work, the starting point of the analysis was a doubling of the fields, a method which has been naturally used for the corresponding quantum problem  where separate fields are needed for time evolution in the direct amplitude and the complex conjugate amplitude. However, when occupation numbers are large one combination of the fields, $\pi$ in \cite{Mueller:2002gd}, becomes a variable of constraint and the functional integration over $\pi$ requires the other independent combination of fields, $\phi$, to obey the classical equations of motion of the $\lambda \phi^4$ theory. Thus, although there is only one dynamical variable in the discussion given in \cite{Mueller:2002gd}, the constraint variable appears explicitly in the perturbative classical calculation of the Boltzmann collision term. 

There are two major differences between the current work and the one in \cite{Mueller:2002gd}. The first is that we simply use a different method which does not rely on the doubling of the fields; we solve classical equations of motion, with retarded boundary conditions as appropriate to the problem, in which only one field evolves and interacts. Occupation numbers are not defined in terms of Green's function, as usually done in the quantum analyses and in that of \cite{Mueller:2002gd}. Instead we start from the ``canonical'' definition that $f_{\bp}$ should be proportional to $a^*_{\bp} a_{\bp}$ where $a^*_{\bp}$ and $a_{\bp}$ are the classical analogues of creation and annihilation operators, i.e.~the coefficients in the expansion of the classical field in plane waves. In this language it is clear how the constraint $f_{\bp} \gg 1$ emerges, since in the classical treatment we consider these expansion coefficients as numbers and not as operators, thus effectively ignoring all possible commutators. Now we can follow the classical time evolution of the field coefficients and in turn that of the occupation numbers.

The second difference with respect to~\cite{Mueller:2002gd}, is that we extend the analysis to the case of a Yang-Mills theory. In order to efficiently deal with the latter, we shall first consider a scalar theory with both cubic, $g \phi^3$, and quartic, $\lambda \phi^4$, interactions. Then the study of the Yang-Mills theory becomes much easier since the topology in the diagrammatic expansion is the same with the only additional complication being the introduction of spin and color degrees of freedom. Our calculations, using classical field equations as already stressed, are given as the first terms in a power series in $g^2$ and $\lambda$ in the scalar theory and in $g^2$ in the Yang-Mills theory. They agree with the corresponding quantum field theory result so long as occupation numbers satisfy $f_{\bp} \gg 1$ and after ensemble averages (whose particular details should not matter when the constraints in the occupation numbers are satisfied)  over the initial conditions are performed in both the classical and quantum approaches. Thus, we shall eventually arrive at the collision integral in \eqn{cf}, but it will contain only the cubic in $f$ terms and not the quadratic ones, cf.~\eqn{beym}. The equilibrium limit in that equation is now given by $f_{\bp}=k T/E_{\bp}$, which is clearly the large occupation limit of the Bose-Einstein distribution occurring when $E_{\bp} \ll kT$.

In order to make our discussion as simple as possible we have made a number of assumptions: (i) We suppose that the elements of our initial ensemble of field configurations are homogeneous in space. This assumption is not really necessary, but it is simplifies our task considerably. What one must actually assume is that inhomogeneities occur on a scale large compared to the wavelengths dominating the problem and this is sufficient to get an effective momentum conservation, e.g.~the $\delta^{(3)}(\Delta \bp)$ emerging in \eqn{deltadp}. When such spatial inhomogeneities are present they trivially give rise to the drift term $\bm{v}_{\bp} \cdot \del f_{\bp}/\del \bx$ which appears in the Boltzmann equation in \eqn{beintro} and combines with $\del f_{\bp}/\del t$ term to form the natural ``convective'' derivative. (ii) We assume the absence of long range coherent fields which would give rise to the term $\bm{F}_{\rm ext}\cdot\del f_{\bp}/\del\bp$ in \eqn{beintro}. (iii) We finally suppose that our initial fields ensemble does not have long range coherences in wavelengths so that \eqn{fpdef} which defines the occupation numbers is appropriate. Similar assumptions were made in the analysis of \cite{Mueller:2002gd}, however, other possibilities are available as we now discuss.

The above assumptions are generally satisfied in recent studies of scalar field theories and their simulations \cite{Blaizot:2011xf,Epelbaum:2011pc,Epelbaum:2014yja}. However, in simulations of Yang-Mills theories this is not always the case.  On the one hand, in \cite{Berges:2013fga,Berges:2013lsa} the initial conditions are very much as we have taken them and one expects that after a short time, allowing occupation numbers to become less than $1/\alpha_s$, the classical field theory simulations should agree with the Boltzmann equation. And indeed, this seems to be the case as the results in \cite{Berges:2013fga,Berges:2013lsa} are very close to the Boltzmann based description given in \cite{Baier:2000sb}. On the other hand, the recent simulations in \cite{Gelis:2013rba} begin with long range coherent fields and thus \eqn{fpdef} is not satisfied. At this point it is not clear at what time the classical field evolution of \cite{Gelis:2013rba} would admit an equivalent description via a Boltzmann equation.

In Sect.~\ref{sec:scalar} we do the derivation for the scalar theory with $g\phi^3$ and $\lambda \phi^4$ interactions. The calculation is based on suitable Feynman rules which allow for a diagrammatic solution of the classical equations of motion. We have separated the calculation in three subsections in which we calculate in great detail the $\lambda^2$, the $\lambda g^2$ and the $g^4$ terms respectively. Each of the aforementioned terms contains all the gain and loss terms of the collision integral.  Then, in Sect.~\ref{sec:ym} we give the derivation for a Yang-Mills theory by paying special attention to the points that require extra treatment compared to the scalar theory case.

%%%%%%%%%%%%%%%%%%%%%%%%%%%%%%%%%%%%%%%%%%%%%%%%%%%%%%%
%%%%%%%%%%%%%%%%%%%%%%% SCALAR %%%%%%%%%%%%%%%%%%%%%%%%
%%%%%%%%%%%%%%%%%%%%%%%%%%%%%%%%%%%%%%%%%%%%%%%%%%%%%%%

\section{\label{sec:scalar} Scalar field theory with cubic and quartic vertices}

Let us start by considering a massless scalar field theory with cubic and quartic interactions in $D=4$ dimensions. The action is given by
\beq
S_{\phi} = \int \dif^4 x \,\mcal{L}_{\phi} = \int \dif^4 x 
\left[ \frac{1}{2} (\del_{\mu} \phi)^2 - 
\frac{g}{3!}\, \phi^3 - \frac{\lambda}{4!}\, \phi^4\right],
\eeq
and while the coupling $\lambda$ is dimensionless, the coupling $g$ has mass dimension 1. In this work, and in view of the perturbation theory to follow, we shall assume that $\lambda$ and $g^2/M^2$ are of the same order, where $M$ is a typical mass scale for the scattering processes to be taken into account. In general, we can decompose the real classical field $\phi$ according to
\beq
\label{phiexp}
\phi(x) = \int \frac{\dif^3 \bp}{h_{\bp}}\, 
\left( a_{\bp}\, \rme^{-\rmi p\cdot x}
+ a_{\bp}^*\, \rme^{\rmi p \cdot x}\right ) \quad \mathrm{with} \quad h_{\bp} = \sqrt{(2\pi)^3 2 E_{\bp}}\,,
\eeq 
and where $p$ is an on-shell four-momentum so that $p\cdot x = E_{\bp}\, x^0 - \bp \cdot \bx$ and $E_{\bp} = |\bp|$. Since we have an interacting field theory, the coefficients $a_{\bp}$ and $a_{\bp}^*$ are generally time-dependent. However, the Boltzmann equation is valid when the typical collision time is much smaller than the time between two collisions. Thus, even though we will assume that $a_{\bp}$ is time-dependent, we will take this dependence to be much slower than that of the plane wave in \eqn{phiexp}. This allows us to invert \eqn{phiexp} and express $a_{\bp}$ in terms of the field 
$\phi$ as
\beq
\label{ap}
a_{\bp} = \frac{\rmi}{h_{\bp}}
\int \dif^3\bx \, \rme^{\rmi p \cdot x} \big[\dot{\phi}(x) - \rmi E_{\bp} \phi(x) \big].  
\eeq

In the case of a homogeneous medium it is natural to define the occupation number $f_{\bp}$, a dimensionless quantity, as
\beq
\label{fpdef}
\left\langle a^*_{\bp'} a_{\bp} \right\rangle = \delta^{(3)}_{\bp\bp'}\, f_{\bp},
\eeq
with the shorthand notation $\delta^{(3)}_{\bp\bp'} \equiv \delta^{(3)}(\bp-\bp')$ and where the brackets stand for the ensemble average. We aim to find the time evolution of the occupation number in the classical theory and therefore we need to determine the corresponding evolution of the coefficients $a_{\bp}$ and the field $\phi$. The classical equation of motion of 
$\phi$ clearly reads
\beq
\label{ceom}
\Box_{x} \phi = J(x) 
\equiv -\frac{g}{2!}\,\phi^2 - \frac{\lambda}{3!}\,\phi^3,     
\eeq
with the convention $\Box_x = \del_{0}^2 - \nabla^2_{\bx}$ and where we have defined for our convenience the ``current'' $J$. Let us now split the full interacting field $\phi$ according to
\beq
\label{phisplit}
\phi = \phi^{(0)} + \delta \phi, 
\eeq
where $\phi^{(0)}$ is the free field, i.e.~it satisfies the homogeneous version \eqn{ceom}, while $\delta\phi$ is the modification arising from the presence of interactions, satisfies \eqn{ceom} and thus can be formally written as
\beq
\label{deltaphi}
\delta\phi(x) = \int \dif^4 y\, \rmi\Delta(x-y) J(y).
\eeq
In the above $\Delta$ is the free propagator of the scalar field and is determined by
\beq
\Box_x \Delta(x-y) = - \rmi\, \delta^{(4)}(x-y) .
\eeq
The solution to the above is
\beq
\label{prop}
\Delta_{\mathrm R}(x) = 
\int \frac{\dif^4 k}{(2 \pi)^4}\,
\rme^{-\rmi k \cdot x} \Delta_{\mathrm R}(k)
=
\int \frac{\dif^4 k}{(2 \pi)^4}\,
\rme^{-\rmi k \cdot x}\,
\frac{\rmi}{k^2 + \rmi \epsilon k^0},
\eeq
where $\epsilon \to 0^+$ so that the propagator is proportional to $\Theta(x^0)$ as it is straightforward to check by performing the integration over $k^0$. More precisely, one finds
 \beq
 \label{proptheta}
 \Delta_{\rm R}(x) =
 -\rmi \Theta(x^0)
 \int \frac{\dif^3 \bk}{(2\pi)^3}\,
 \frac{\sin(E_{\bk}\, x^0)}{E_{\bk}}\,
 \rme^{\rmi \bk \cdot \bx}.
 \eeq 
Therefore, the propagator in \eqn{prop} is the retarded (or causal) one, since this is the natural choice when initial conditions (that is, $\phi^{(0)}$) are given. For later use let us note that this retarded propagator can  also be written as
\beq
\label{propsplit}
\Delta_{\mathrm R}(k) = \frac{\rmi}{2 E_{\bk}}
\left[\frac{\mathrm{P}}{k^0 - E_{\bk}} - \rmi \pi \delta(k^0 - E_{\bk}) 
-\frac{\mathrm{P}}{k^0 + E_{\bk}} + \rmi \pi \delta(k^0 + E_{\bk}) 
\right],
\eeq
where P stands for principal value and therefore one has a clear separation of the real and imaginary contributions to the propagator. 

Now, in analogy to \eqn{phisplit} we can split the coefficient $a_{\bp}$ as
\beq
a_{\bp} = a^{(0)}_{\bp} + \delta a_{\bp}
\eeq
and using the form of the propagator given in \eqn{proptheta} just above we easily find that the piece $\delta a_{\bp}$ generated by the interactions is given by
\beq
\label{deltaap}
\delta a_{\bp} = \frac{\rmi}{h_{\bp}}
\int \dif^4 y \, \rme^{\rmi p \cdot y}\,
\Theta(x^0 - y^0) J(y).
\eeq
The corresponding change in the occupation number reads
\beq
\label{deltafp}
\delta^{(3)}_{\bp\bp'}\, \delta f_{\bp} =
2 \mathrm{Re} 
\big[\big\langle a^{(0)*}_{\bp'} \delta a_{\bp} \big\rangle\big]
+\big\langle\delta a^*_{\bp'}\, \delta a_{\bp}\big\rangle, 
\eeq 
where, in writing the first term on the r.h.s.~of the above, we have anticipated that it will be proportional to $\delta^{(3)}_{\bp\bp'}$ like the l.h.s. Finally, by taking a time derivative we arrive at
\beq
\label{fpdot}
\delta^{(3)}_{\bp\bp'}\, \dot{f}_{\bp}
= 2 \mathrm{Re} 
\big[\big\langle a^{(0)*}_{\bp'} \delta \dot{a}_{\bp} \big\rangle\big]
+2 \mathrm{Re} 
\big[\big\langle \delta a^*_{\bp'}\, \delta \dot{a}_{\bp} \big\rangle\big],
\eeq
where, with a slight notational abuse, $\delta \dot{a}_{\bp}$ stands for the time derivative of $\delta a_{\bp}$. We shall refer to the two terms in the r.h.s.~of \eqn{fpdot} as the crossed and diagonal terms respectively.

In general one cannot solve \eqn{deltaphi} and/or \eqn{deltaap}; that would be equivalent to solving the full nonlinear classical problem, which is in any case beyond our goals. What we shall do, is to assume that the correction $\delta a_{\bp}$ is small compared to $a_{\bp}^{(0)}$ and perform a calculation to first non-vanishing order in $\lambda \sim g^2/M^2$. Eventually this translates to imposing that occupations numbers do not get large, more precisely $f_{\bp} \ll 1/\lambda$. Recalling that the classical approximation to the problem also requires $f_{\bp} \gg 1$, we see that there is a parametrically large window of validity for the ``classical'' Boltzmann equation, so long as the couplings are sufficiently small.

\subsection{\label{subsec:l2} The $\lambda^2$ terms and the Feynman rules for classical diagrams in the scalar theory}

To illustrate the procedure, we shall first do a step-by-step calculation for the $\lambda^2$ contribution to the diagonal term in \eqn{fpdot} which simply means that we need to find the order $\lambda$ contribution to $\delta a_{\bp}$. Since the current in \eqn{ceom} is already of order $\lambda$ we can substitute the full field $\phi$ with its free part $\phi^{(0)}$. Next, for reasons to be apparent in a while, let us consider the following particular term in $[\phi^{(0)}(y)]^3$
\beq
\label{phi03}
[\phi^{(0)}(y)]^3 \to 3
\int \frac{\dif^3 \bp_1}{h_{\bp_1}}\,
\frac{\dif^3 \bp_2}{h_{\bp_2}}\,
\frac{\dif^3 \bp_3}{h_{\bp_3}}\,
a^{(0)*}_{\bp_1} a^{(0)}_{\bp_2} a^{(0)}_{\bp_3}
\rme^{\rmi (p_1 - p_2 -p_3)\cdot y},
\eeq
where $p_1$, $p_2$ and $p_3$ are on-shell four-momenta and with the combinatorial factor 3 coming from the number of ways we can pick the required product of field coefficients out of $[\phi^{(0)}]^3$. Now we can integrate over $\by$ to get
\beq
\label{deltadp}
\delta a_{\bp} = -\frac{\rmi}{h_{\bp}}\,\frac{\lambda}{2}
\int \frac{\dif^3 \bp_1}{h_{\bp_1}}\,
\frac{\dif^3 \bp_2}{h_{\bp_2}}\,
\frac{\dif^3 \bp_3}{h_{\bp_3}}\,
(2\pi)^3 \delta^{(3)}(\Delta \bp)\,
a^*_{\bp_1} a_{\bp_2} a_{\bp_3}
\int \dif y_0\, \Theta(x^0-y^0)
\rme^{\rmi \Delta E y_0}  
\eeq
where we have defined $\Delta \bp = \bp + \bp_1 - \bp_2 -\bp_3$ and 
$\Delta E = E_{\bp}+ E_{\bp_1} - E_{\bp_2} - E_{\bp_3}$. Notice that we have dropped the superscript $(0)$ from the expansion coefficients, since this is allowed to the level of accuracy and in order to have a more economical notation. Furthermore, let us point out that at this stage energy is not conserved at the vertex. The $y^0$ time integration is unbounded for large negative values and we make it convergent via the ``adiabatic'' prescription $\Delta E \to \Delta E - \rmi \epsilon$
with $\epsilon \to 0^+$ to find
\beq
\label{dapltemp}
\delta a_{\bp} = \frac{1}{h_{\bp}}\,\frac{(-\rmi\lambda)}{2} 
\int \frac{\dif^3 \bp_1}{h_{\bp_1}}\,
\frac{\dif^3 \bp_2}{h_{\bp_2}}\,
\frac{\dif^3 \bp_3}{h_{\bp_3}}\,
(2\pi)^3 \delta^{(3)}(\Delta \bp)\,
\frac{\rme^{\rmi (\Delta E - \rmi \epsilon) x^0}}{\rmi(\Delta E - \rmi \epsilon)}\,
a^*_{\bp_1} a_{\bp_2} a_{\bp_3}.
\eeq
From the above ``direct amplitude'' (DA) it is straightforward to construct its time derivative $\delta \dot{a}_{\bp}$ and the ``complex conjugate amplitude'' (CCA) $\delta a_{\bp'}^*$. When forming $\langle \delta a_{\bp'}^* \delta \dot{a}_{\bp} \rangle$ we encounter a six-point correlator of the field coefficients and since the system is dilute we will assume that it factorizes to a product of two-point functions, that is, to a product of occupation numbers. More precisely, we assume the ensemble average
\begin{align}
\label{fact}
\big\langle a^*_{\bp_1} a_{\bp_2} a_{\bp_3} 
a_{\bp'_1} a^*_{\bp'_2} a^*_{\bp'_3}
\big\rangle
&\to
\big\langle a^*_{\bp_1} a_{\bp'_1} \big\rangle 
\big\langle a^*_{\bp_2} a_{\bp'_2} \big\rangle 
\big\langle a^*_{\bp_3} a_{\bp'_3} \big\rangle 
+
\big\langle a^*_{\bp_1} a_{\bp'_1} \big\rangle 
\big\langle a^*_{\bp_2} a_{\bp'_3} \big\rangle 
\big\langle a^*_{\bp_3} a_{\bp'_2} \big\rangle  
\nn
&= \Big[
\delta^{(3)}_{\bp_1\bp'_1}
\delta^{(3)}_{\bp_2\bp'_2} 
\delta^{(3)}_{\bp_3\bp'_3}
+
\delta^{(3)}_{\bp_1\bp'_1}
\delta^{(3)}_{\bp_2\bp'_3} 
\delta^{(3)}_{\bp_3\bp'_2}\Big]
f_{\bp_1}f_{\bp_2}f_{\bp_3} 
\end{align}
and since we integrate over all momenta one immediately sees that both terms in the above will eventually contribute the same to the final result. Using the $\delta$-functions arising from the ensemble average in \eqn{fact} one can readily perform all the integrations over the primed momenta in the product $\langle \delta a_{\bp'}^* \delta \dot{a}_{\bp} \rangle$. Then the $\delta$-function corresponding to momentum conservation in the CCA becomes $\delta^{(3)}(\bp'+\bp_1 - \bp_2 - \bp_3)$ and after also using momentum conservation in the DA it finally gives a factor $\delta^{(3)}_{\bp\bp'}$ as expected (cf.~the discussion after \eqn{deltafp}). Now $\Delta E$ becomes the same in the DA and in the CCA and we have
\beq
\label{rede}
\mathrm{Re}\, \frac{\rmi}{\Delta E + \rmi \epsilon}
= \frac{\epsilon}{(\Delta E)^2 + \epsilon^2} = 
\pi \delta(\Delta E),
\eeq
which is the required energy conservation. Now we put everything together in \eqn{fpdot} to finally arrive at the $\lambda^2$ gain term
\beq
\label{gainl2}
\dot{f}_{\bp}\big|_{\lambda^2}^{A} = \frac{1}{4 E_{\bp}} 
\int \widetilde{\dif \bp_1}\,
\widetilde{\dif \bp_2}\,
\widetilde{\dif \bp_3}\,
(2\pi)^4 \delta^{(4)}(\Delta p)\,\lambda^2
f_{\bp_1}f_{\bp_2}f_{\bp_3},
\eeq
where $\Delta p = p + p_1 - p_2 - p_3$ with all four-momenta being on-shell and where we have adopted the compact notation introduced in \eqn{dptilde} for the integration measure.

\begin{figure}
\begin{minipage}[b]{0.25\textwidth}
\includegraphics[scale=0.50]{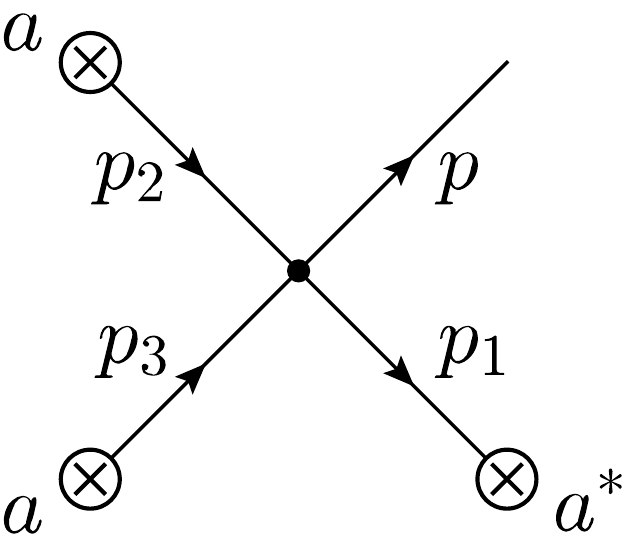}
\end{minipage}
\begin{minipage}[b]{0.05\textwidth}
$\mbox {\LARGE $\times$}$
\vspace{1.2cm}
\end{minipage}
\begin{minipage}[b]{0.25\textwidth}
\includegraphics[scale=0.50]{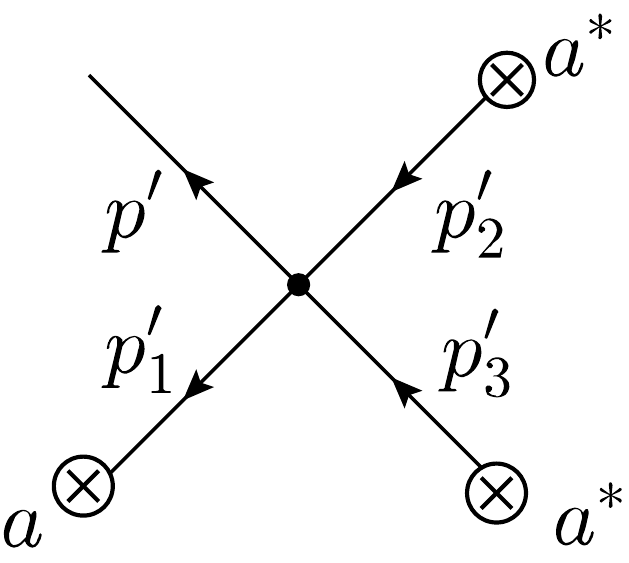}
\end{minipage}
\caption{The $\lambda^2$ contribution to $\delta a_{\bp} \delta a^*_{\bp'}$, cf.~\eqn{dapltemp}. A circled cross stands for an external insertion while the open line corresponds to the momentum measured. The ensemble average will set $\bp_1'=\bp_1$, $\bp_2'=\bp_2$  and $\bp_3'=\bp_3$ (or $\bp_2'=\bp_3$ and $\bp_3'=\bp_2$), while momentum conservation in both the DA and in the CCA will lead to $\bp'=\bp$.}
\label{fig:dapdapl2}
\end{figure}

Let us note here that it is only the choice made in \eqn{phi03} for the field coefficients which leads to energy conservation. Any other combination, e.g.~an $a^*a^*a$ term, will lead to complex exponentials with uncompensated energy differences. Such exponentials will average to zero at large times, since the time scales describing variations in the Boltzmann equation are supposed to be very large compared to the typical interaction times. $\lambda^2$ is simply the amplitude squared $|\mcal{M}(p_2 p_3;p p_1 )|^2$ in the $\lambda \phi^4$ theory and \eqn{gainl2} acquires a natural interpretation as a gain term arising from a $2 \to 2$ scattering. The integrand is naturally proportional to the occupation numbers of the incoming particles $f_{\bp_2}$ and $f_{\bp_3}$, while $f_{\bp_1}$ appears as a Bose enhancement factor. The (square of the) Feynman diagram related to the term we have just calculated is shown in Fig.~\ref{fig:dapdapl2}.
    
Let us now establish some Feynman rules for the classical problem at hand in order to systematize the calculation for the remaining terms. For any diagram in the DA we have the following momentum space rules
\begin{list}{\small $\Box$}{\setlength{\itemsep}{0cm} \setlength{\itemindent}{0cm}}
\item Assign a factor $1/h_{\bp}$ from the definition of $\delta a_{\bp}$.
\item Assign a factor $-\rmi g$ for each cubic vertex and a factor $-\rmi \lambda$ for each quartic one.
\item Divide by the symmetry factor. The maximum such factor we will come across is 2; this will take place when two field coefficients of the same type, that is two $a$'s or two $a^*$'s, are connected to the same vertex.
\item Impose three-momentum conservation at each vertex.
\item Assign an overall factor $(2\pi)^3 \delta^{(3)}(\overline{\Delta \bp})$ where $\overline{\Delta \bp}$ is the sum over all external three-momenta in which the momentum $\bp$ and the momenta associated with $a^*$'s are taken with a positive sign, while the momenta associated with $a$'s are taken with a negative sign. 
\item Impose energy conservation at all, but one (see next rule), vertices.
\item Assign a factor $\exp[\rmi(\overline{\Delta E}-\rmi \epsilon)x^0]/[\rmi (\overline{\Delta E}-\rmi \epsilon)]$ with $\epsilon \to 0^+$ at the vertex which connects to the measured occupation factor. $\overline{\Delta E}$ is the energy imbalance at the vertex, and thus also that of the full diagram, with $E_{\bp}$ taken with a positive sign.
\item Use the retarded propagator $\Delta_{\mathrm R}(k) = \rmi/(k^2 + \rmi \epsilon k^0)$, with $\epsilon\to 0^+$, for each internal line. The four-momentum $k$ should flow towards the measured occupation factor. Equivalently, one can use the advanced propagator $\Delta_{\mathrm A}(k) = \Delta_{\mathrm R}(-k) = \rmi/(k^2 - \rmi \epsilon k^0)$ if the four-momentum $k$ is taken to flow away from the measured occupation factor.
\item Integrate according to $\displaystyle{\int \frac{\dif^3 \bp}{h_{\bp}}}\, a^*_{\bp}$ or
$\displaystyle{\int \frac{\dif^3 \bp}{h_{\bp}}}\, a_{\bp}$ for each external line, but not for the measured particle. 
\end{list}
We stress that these rules are just a convenient representation of the perturbative solution to the classical problem. It is trivial to check that they lead to \eqn{dapltemp} when considering the DA in Fig.~\ref{fig:dapdapl2}.

\begin{figure}
\begin{minipage}[b]{0.45\textwidth}
\begin{center}
\includegraphics[scale=0.50]{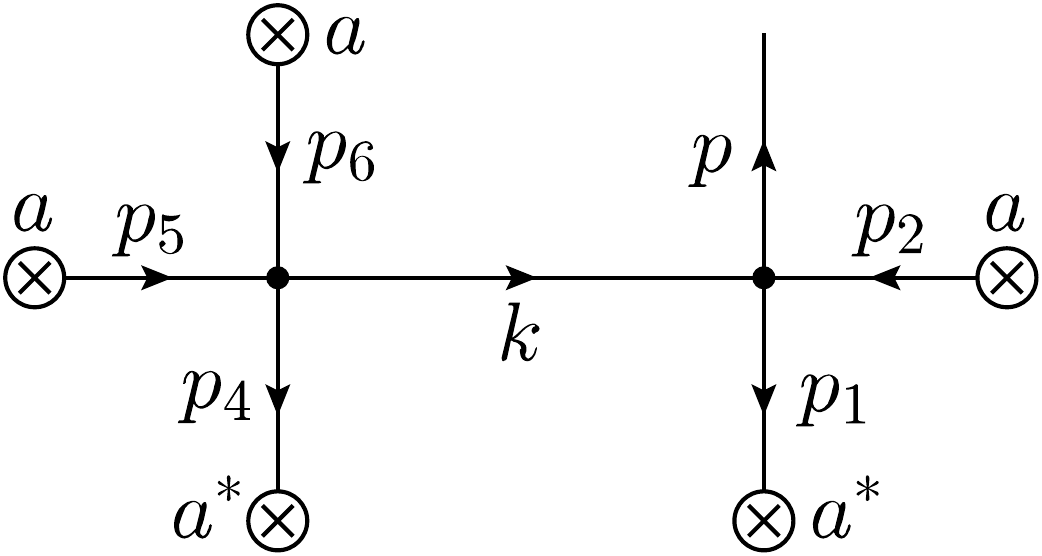}\\{\small (a)}
\end{center}
\end{minipage}
\begin{minipage}[b]{0.45\textwidth}
\begin{center}
\includegraphics[scale=0.50]{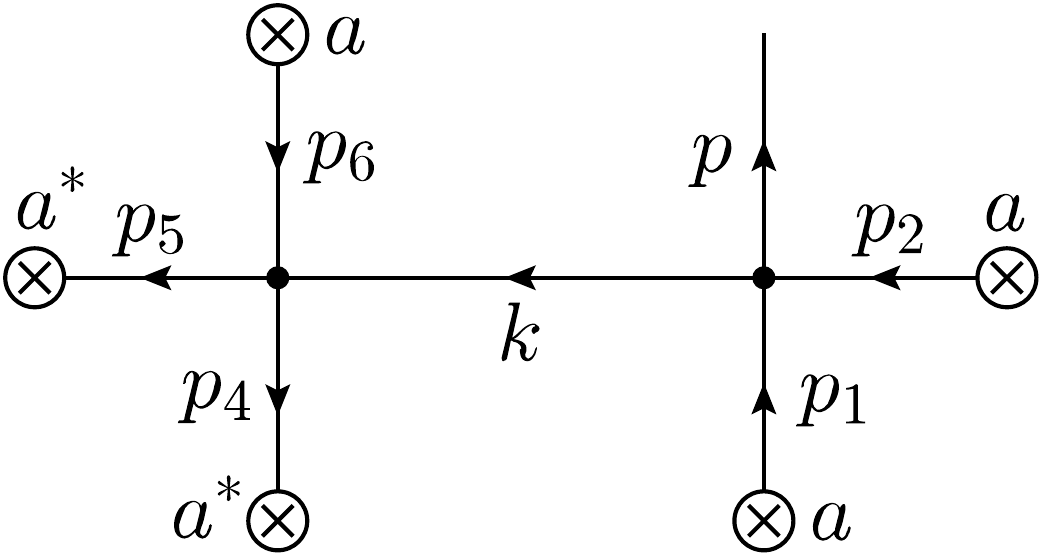}\\{\small (b)}
\end{center}
\end{minipage}
\caption{The $\lambda^2$ contributions to $\delta a_{\bp}$ leading to (a) the loss terms and (b) a gain term in the Boltzmann equation.}
\label{fig:dapl2}
\end{figure}

Next, we shall use these Feynman rules to calculate the remaining $\lambda^2$ terms. These come from the crossed term in \eqn{fpdot} and it is clear that now we need to compute $\delta a_{\bp}$ to order $\lambda^2$. To this order, the two diagrams which will eventually satisfy energy conservation are shown in Fig.~\ref{fig:dapl2}. As we shall see, the diagram \ref{fig:dapl2}.a leads to the loss terms in the Boltzmann equation while \ref{fig:dapl2}.b leads to a gain term.

Even though it is not necessary, let us say, just for illustrative purposes, that such diagrams arise from the current $J(y)$ expanded to order $\lambda^2$ which can be easily found to be
\beq
J(y) = -\frac{\lambda^2}{12}\, \phi^2(y)
\int \dif^4 z\, \rmi \Delta(y-z) \phi^3(z),
\eeq  
where we have dropped the superscript $(0)$ in the field $\phi$. Now one would need to expand all the free fields in plane waves as before, but as explained above it is more convenient and much less tedious to directly use the Feynman rules. We readily see that the diagram \ref{fig:dapl2}.a gives
\beq
\label{dapl2atemp}
\delta a_{\bp} = - 
\frac{1}{h_{\bp}}\,\frac{\lambda^2}{2}
\int \prod_{i} \frac{\dif^3 \bp_{i}}{h_{\bp_i}}\,
(2\pi)^3 \delta^{(3)}(\overline{\Delta\bp})\,
\frac{\rme^{\rmi (\overline{\Delta E} - \rmi \epsilon) x^0}}{\rmi(\overline{\Delta E} - \rmi \epsilon)}\,
\Delta_{\mathrm R}(k)\,
a^*_{\bp_1} a_{\bp_2} a^*_{\bp_4} a_{\bp_5} a_{\bp_6},
\eeq
with $\overline{\Delta\bp} = \bp + \bp_1 +\bp_4 - \bp_2 -\bp_5 -\bp_6$,  $\overline{\Delta E} = E_{\bp} + E_{\bp_1} +E_{\bp_4} - E_{\bp_2} -E_{\bp_5} -E_{\bp_6}$ and $k = p_5 + p_6 - p_4$. The symmetry factor 2 in the denominator comes about because the diagram remains invariant under the exchange of the legs corresponding to momenta $p_5$ and $p_6$. Differentiation w.r.t~$x^0$ cancels the energy denominator and multiplication with 
$a^*_{\bp'}$ (cf.~\eqn{fpdot}) leads again to a product of six field coefficients. As in \eqn{fact} we assume that the six-point correlator factorizes into a product of occupation numbers, that is,
\beq
\label{fact2}
\big \langle 
a^*_{\bp_1} a_{\bp_2} a^*_{\bp_4} a_{\bp_5} a_{\bp_6} a^*_{\bp'}
\big\rangle
\to 2 \delta^{(3)}_{\bp_1\bp_5}
\delta^{(3)}_{\bp_2\bp_4} 
\delta^{(3)}_{\bp_6\bp'}
f_{\bp'}f_{\bp_1} f_{\bp_2}.
\eeq
The factor of 2 comes because $\bp_5$ has to be contracted with either $\bp_1$ or $\bp'$ (and, correspondingly, $\bp_6$ with either $\bp'$ or $\bp_1$) and both terms contribute equally. The $\delta$-function in the integrand of \eqn{dapl2atemp} reduces to  $\delta^{(3)}_{\bp\bp'}$, and then $\overline{\Delta E}$ vanishes and $k$ becomes $p + p_1 - p_2$. Furthermore, making use of \eqn{propsplit} we have
\beq
\label{repropr}
\mathrm{Re}
\left(-\frac{\rmi}{k^2 + \rmi \epsilon k^0}\right)
=\mathrm{Im}
\left(\frac{1}{k^2 + \rmi \epsilon k^0}\right)
= -\frac{\pi}{2 E_{\bk}}\,\delta(k^0 - E_{\bk}), 
\eeq
which expresses energy conservation. Notice that due to the three $\delta$-functions in \eqn{fact2}, there are only two three-momentum integrations to be done which means the $\delta$-function of the three-momentum conservation has been already implicitly used. To comply with the notation of \eqn{gainl2} one can re-insert an integration over the momentum $\bk$, which we rename to $\bp_3$, accompanied by $\delta^{(3)}(\bp + \bp_1 - \bp_2 - \bp_3)$. Then by putting everything together in \eqn{fpdot} we arrive at the order $\lambda^2$ loss terms
\beq
\label{lossl2}
\dot{f}_{\bp}\big|_{\lambda^2}^{B} = - \frac{1}{4 E_{\bp}} 
\int \widetilde{\dif \bp_1}\,
\widetilde{\dif \bp_2}\,
\widetilde{\dif \bp_3}\,
(2\pi)^4 \delta^{(4)}(\Delta p)\,\lambda^2
\big[ f_{\bp}f_{\bp_1}f_{\bp_2} +
f_{\bp}f_{\bp_1}f_{\bp_3} \big],
\eeq
where, as in \eqn{gainl2}, $\Delta p = p+ p_1 -p_2 -p_3$ with all four-momenta on-shell. Notice that we have been allowed to let $2 f_{\bp}f_{\bp_1}f_{\bp_2}$ $\to$ $f_{\bp}f_{\bp_1}f_{\bp_2} + f_{\bp}f_{\bp_1}f_{\bp_3}$ in the integrand in \eqn{lossl2}. Even though diagram \ref{fig:dapl2}.a does not initially look like $2\to 2$ scattering, such an interpretation is eventually possible since the propagator $\Delta_{\rmR}(k)$ is put on-shell (cf.~\eqn{repropr}). Thus, the diagram \ref{fig:dapl2}.a does look like the amplitude squared $|\mcal{M}|^2$ for $2 \to 2$ scattering. Indeed, this is apparent in the loss terms of the Boltzmann equation given in \eqn{lossl2}; the integrand is proportional to $\lambda^2$ and to the occupation numbers $f_{\bp}$ and $f_{\bp_1}$ of the ``incoming'' momenta while $f_{\bp_2}$ ($f_{\bp_3}$) in the first (second) term is a Bose enhancement factor.

Similarly, the diagram \ref{fig:dapl2}.b gives
\beq
\label{dapl2btemp}
\delta a_{\bp} = - 
\frac{1}{h_{\bp}}\,\frac{\lambda^2}{4}
\int \prod_{i} \frac{\dif^3 \bp_{i}}{h_{\bp_i}}\,
(2\pi)^3 \delta^{(3)}(\overline{\Delta\bp})\,
\frac{\rme^{\rmi (\overline{\Delta E} - \rmi \epsilon) x^0}}{\rmi(\overline{\Delta E} - \rmi \epsilon)}\,
\Delta_{\rm A}(k)\,
a_{\bp_1} a_{\bp_2} a^*_{\bp_4} a^*_{\bp_5} a_{\bp_6},
\eeq
with $\overline{\Delta\bp} = \bp + \bp_4 +\bp_5 - \bp_1 -\bp_2 -\bp_6$,  $\overline{\Delta E} = E_{\bp} + E_{\bp_4} +E_{\bp_5} - E_{\bp_1} -E_{\bp_2} -E_{\bp_6}$ and $k = p_4 + p_5 - p_6$. The symmetry factor 4 in the denominator comes about because the diagram remains invariant under the exchange of the legs corresponding to momenta $p_1$ and $p_2$ and the exchange of the legs corresponding to $p_4$ and $p_5$. The six-point correlator factorizes into a product of occupation numbers according to
\beq
\label{fact3}
\big \langle 
a_{\bp_1} a_{\bp_2} a^*_{\bp_4} a^*_{\bp_5} a_{\bp_6} a^*_{\bp'}
\big\rangle
\to 2 \delta^{(3)}_{\bp_1\bp_5}
\delta^{(3)}_{\bp_2\bp_4} 
\delta^{(3)}_{\bp_6\bp'}
f_{\bp'}f_{\bp_1} f_{\bp_2},
\eeq
where the factor of 2 arises because one can set $\bp_4=\bp_1$, $\bp_5=\bp_2$ or $\bp_4=\bp_2$, $\bp_5=\bp_1$. The momentum $k$ becomes $p_1 + p_2 -p$. For the propagator, which is advanced since we took the momentum to flow away from the measured occupation factor, we have
\beq
\label{repropa}
\mathrm{Re}
\left(-\frac{\rmi}{k^2 - \rmi \epsilon k^0}\right)
=\mathrm{Im}
\left(\frac{1}{k^2 - \rmi \epsilon k^0}\right)
= \frac{\pi}{2 E_{\bk}}\,\delta(k^0 - E_{\bk}).
\eeq
This is the point where the two diagrams in Fig.~\ref{fig:dapl2} differ from each other. Compared to \eqn{repropr} the sign in \eqn{repropa} has changed and therefore diagram \ref{fig:dapl2}.b leads to a gain term. As before we insert an integration over the momentum $\bk$, which we rename to $\bp_3$, accompanied by $\delta^{(3)}(\bp + \bp_3 - \bp_1 - \bp_2)$ and we immediately let $\bp_1 \leftrightarrow \bp_3$. We put everything together in \eqn{fpdot} to arrive at the order $\lambda^2$ second gain term
\beq
\label{gainl2b}
\dot{f}_{\bp}\big|_{\lambda^2}^{C} = \frac{1}{4 E_{\bp}} 
\int \widetilde{\dif \bp_1}\,
\widetilde{\dif \bp_2}\,
\widetilde{\dif \bp_3}\,
(2\pi)^4 \delta^{(4)}(\Delta p)\,\lambda^2
f_{\bp}f_{\bp_2}f_{\bp_3},
\eeq
where $\Delta p$ is as in Eqs.~\eqref{gainl2} and \eqref{lossl2}. Again, as already explained below \eqn{lossl2} for the corresponding loss term, the diagram \ref{fig:dapl2}.b eventually acquires an interpretation as $2\to 2$ scattering. The integrand is proportional to the scattering amplitude squared $\lambda^2$ and to the occupation numbers $f_{\bp_2}$ and $f_{\bp_3}$ of the ``incoming'' momenta while $f_{\bp}$ is a Bose enhancement factor. 

Adding all the $\lambda^2$ contributions from Eqs.~\eqref{gainl2}, \eqref{lossl2} and \eqref{gainl2b} we arrive in fact at the Boltzmann equation in the classical $\phi^4$ theory, that is
\beq
\label{alll2}
\dot{f}_{\bp}\big|_{\lambda^2} = \frac{1}{4 E_{\bp}} 
\int \widetilde{\dif \bp_1}\,
\widetilde{\dif \bp_2}\,
\widetilde{\dif \bp_3}\,
(2\pi)^4 \delta^{(4)}(\Delta p)\,\lambda^2
\big[
 f_{\bp_2}f_{\bp_3} \big(f_{\bp_1} + f_{\bp}\big)   
- f_{\bp}f_{\bp_1}  \big(f_{\bp_2} + f_{\bp_3}\big) 
\big].
\eeq

\subsection{\label{subsec:g4} The $g^4$ terms}

Let us turn our attention to contributions arising solely from the cubic vertices, i.e.~the $g^4$ terms. What is non-trivial, compared to the $\lambda^2$ terms, is that now the amplitude squared $|\mcal{M}|^2$ depends on the kinematics. This  dependence, containing the well-known $s$, $t$ and $u$ diagrams, should come out from our calculation.

\begin{figure}
\begin{minipage}[b]{0.45\textwidth}
\begin{center}
\includegraphics[scale=0.50]{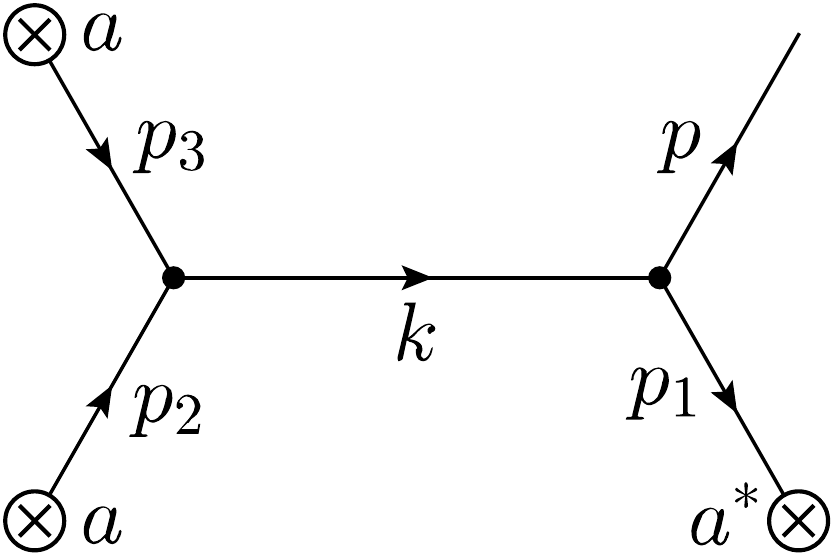}\\{\small (a)}
\end{center}
\end{minipage}
\begin{minipage}[b]{0.45\textwidth}
\begin{center}
\includegraphics[scale=0.50]{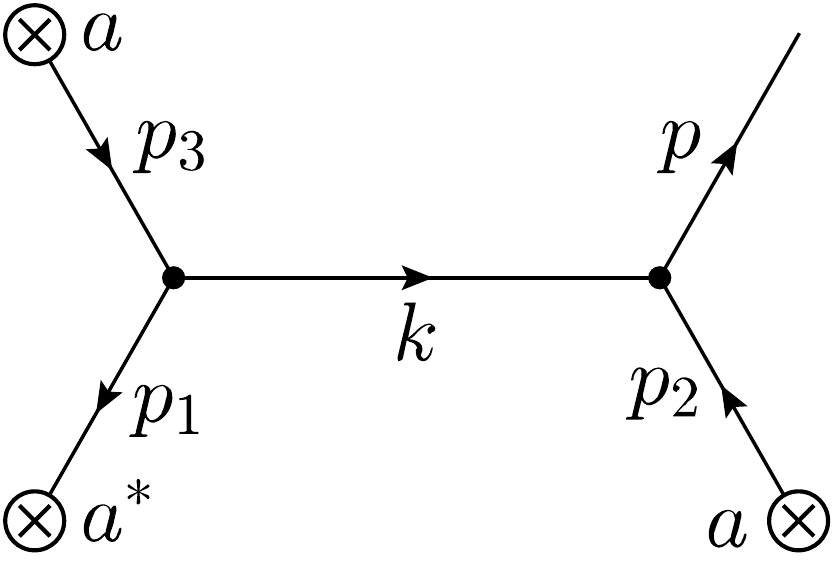}\\{\small (b)}
\end{center}
\end{minipage}
\caption{The $g^2$ contributions to $\delta a_{\bp}$ leading to a gain term in the Boltzmann equation.}
\label{fig:dapg2}
\end{figure}

Before writing down the diagrams, and focusing first on the diagonal term
$\delta a^*_{\bp'}\, \delta \dot{a}_{\bp}$ in \eqn{fpdot}, we give for completeness the current $J(y)$ to order $g^2$; a single iteration leads to
\beq
\label{jg2}
J(y) = -\frac{g^2}{2}\, \phi(y)
\int \dif^4 z\, \rmi\Delta(y-z) \phi^2(z).
\eeq 
The Feynman diagrams for $\delta a_{\bp}$, which in the end will contribute to the Boltzmann equation, are shown in Fig.~\ref{fig:dapg2}. In analogy to the corresponding $\lambda$ term (cf.~\eqn{dapltemp}) we need a product of the type $a^*aa$, and since $a^*$ can originate either from $\phi(y)$ or from $\phi(z)$ we have the two distinct diagrams in Fig.~\ref{fig:dapg2}. Using the Feynman rules we can combine both diagrams into
\beq
\label{dapg2temp}
\delta a_{\bp} = - 
\frac{1}{h_{\bp}}\,\frac{g^2}{2}
\int \prod_{i} \frac{\dif^3 \bp_{i}}{h_{\bp_i}}\,
(2\pi)^3 \delta^{(3)}(\Delta\bp)\,
\frac{\rme^{\rmi (\Delta E - \rmi \epsilon) x^0}}{\rmi(\Delta E - \rmi \epsilon)}\,
\big[\Delta_{\rm R}(p_2 + p_3) + 2
\Delta_{\rm R}(p_3 - p_1)\big]
a^*_{\bp_1} a_{\bp_2} a_{\bp_3},
\eeq
with $\Delta E$ and $\Delta \bp$ as in \eqn{dapltemp}. The two diagrams differ only in the symmetry factors ($1/2$ and $1$ respectively) and in the argument of the retarded propagator. \eqn{dapg2temp} is very similar to \eqn{dapltemp} with the only difference being the presence of a propagator for each of the two terms. In fact, the only role of these propagators is to lead to the proper form of $|\mcal{M}|^2$ in the $g \phi^3$ theory. Therefore the calculation is almost identical to the one following \eqn{dapltemp}. In particular, notice that the real part of the propagators, since they are in general off-shell, does not play any role in the computation of the diagrams under current consideration and the energy conservation will emerge as in \eqn{rede}. We just need to be careful to pick-up the correct arguments of the propagators after the contractions between the DA and the CCA due to the ensemble average. Defining the Mandelstam variables
\beq
\label{stu}
s=(p+p_1)^2, \quad t=(p-p_2)^2 \quad {\rm and} \quad u = (p-p_3)^2 = (p_1 - p_2)^2, 
\eeq 
it is just a matter of simple bookkeeping to find the propagator products after taking the ensemble average of 
$\delta a^*_{\bp'}\, \delta \dot{a}_{\bp}$. For $\bp'_1 = \bp_1$, $\bp'_2 = \bp_2$ and  $\bp'_3 = \bp_3$ (with the prime denoting momenta in the CCA) we have
\begin{align}
&\Delta_{\rm R}(p_2 + p_3)
\Delta^*_{\rm R}(p'_2 + p'_3) 
= 1/s^2,
\nn
&\Delta_{\rm R}(p_2 + p_3)
\Delta^*_{\rm R}(p'_3 - p'_1)
= 1/st,
\nn
&\Delta_{\rm R}(p_3 - p_1)
\Delta^*_{\rm R}(p'_2 + p'_3)
= 1/st,
\nn
&\Delta_{\rm R}(p_3 - p_1)
\Delta^*_{\rm R}(p'_3 - p'_1) 
= 1/t^2,
\end{align}
while for $\bp'_1 = \bp_1$, $\bp'_2 = \bp_3$ and  $\bp'_3 = \bp_2$
\begin{align}
&\Delta_{\rm R}(p_2 + p_3)
\Delta^*_{\rm R}(p'_2 + p'_3) 
= 1/s^2,
\nn
&\Delta_{\rm R}(p_2 + p_3)
\Delta^*_{\rm R}(p'_3 - p'_1)
= 1/su,
\nn
&\Delta_{\rm R}(p_3 - p_1)
\Delta^*_{\rm R}(p'_2 + p'_3)
= 1/st,
\nn
&\Delta_{\rm R}(p_3 - p_1)
\Delta^*_{\rm R}(p'_3 - p'_1) 
= 1/tu.
\end{align}
Putting everything together and noticing that one can let $2/t^2 \to 2/u^2$ and $2/st \to 2/su$ inside the integrand we find the gain term
\beq
\label{gaing4}
\dot{f}_{\bp}\big|_{g^4}^{A} = \frac{1}{4 E_{\bp}} 
\int \widetilde{\dif \bp_1}\,
\widetilde{\dif \bp_2}\,
\widetilde{\dif \bp_3}\,
(2\pi)^4 \delta^{(4)}(\Delta p)\,
\left[\frac{g^2}{s}+\frac{g^2}{t}+\frac{g^2}{u} \right]^2
f_{\bp_1}f_{\bp_2}f_{\bp_3}.
\eeq

\begin{figure}
\begin{minipage}[b]{0.325\textwidth}
\begin{center}
\includegraphics[scale=0.47]{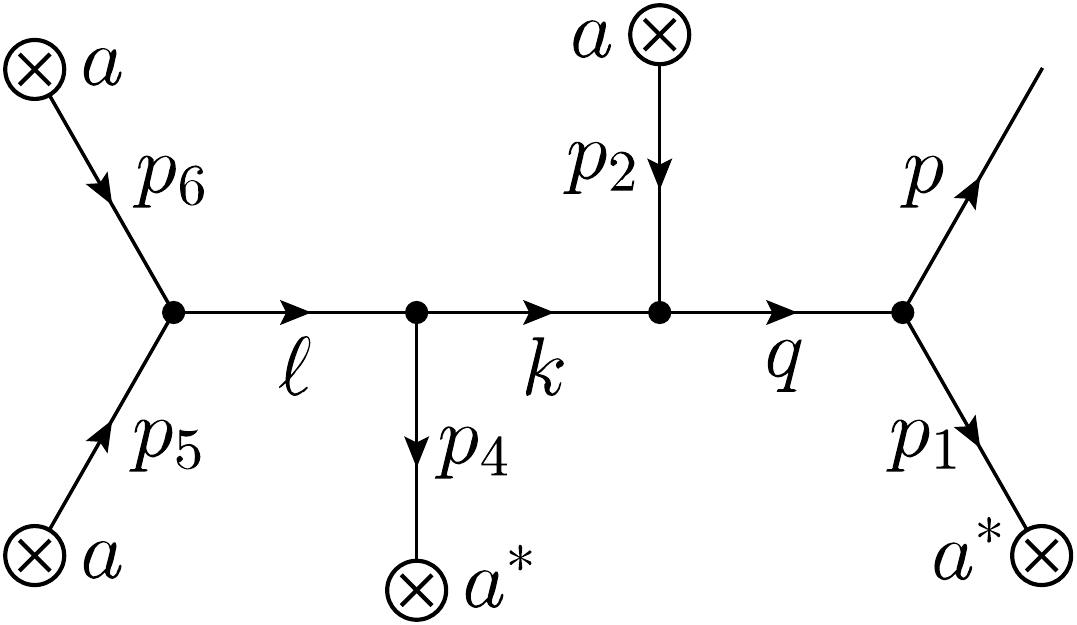}\\{\small (a)}
\end{center}
\end{minipage}
\begin{minipage}[b]{0.325\textwidth}
\begin{center}
\includegraphics[scale=0.47]{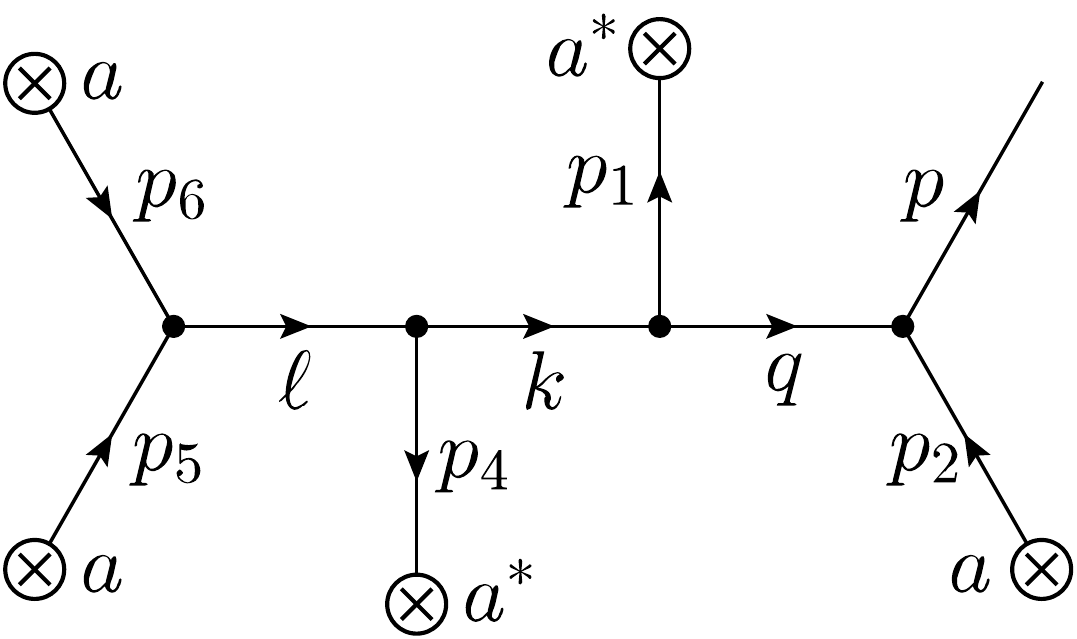}\\{\small (b)}
\end{center}
\end{minipage}
\begin{minipage}[b]{0.325\textwidth}
\begin{center}
\includegraphics[scale=0.47]{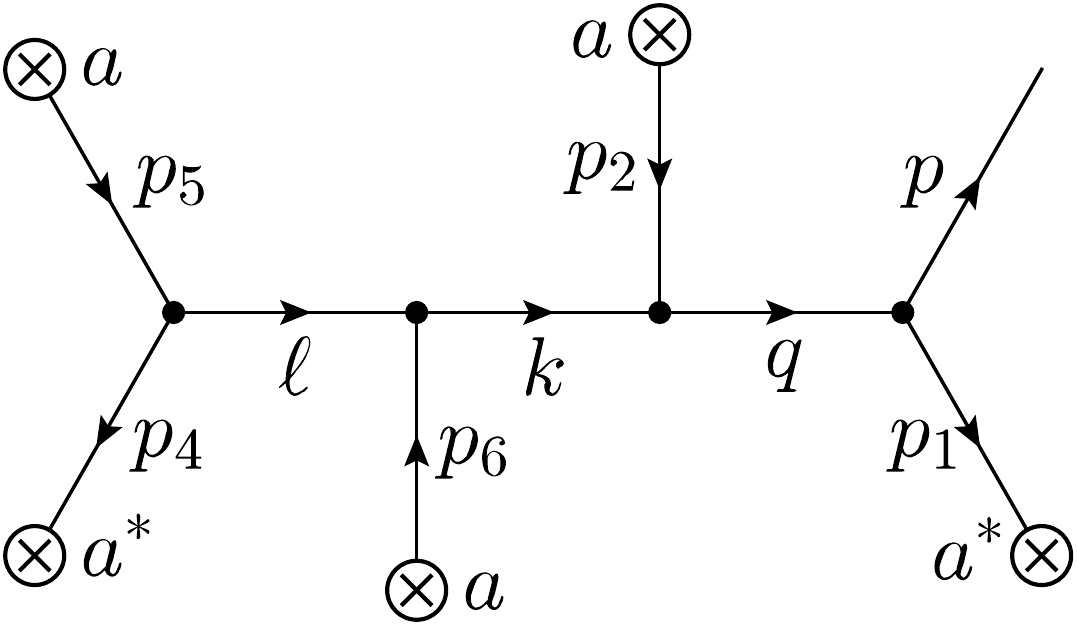}\\{\small (c)}
\end{center}
\end{minipage}
\begin{minipage}[b]{0.325\textwidth}
\vspace{0.5cm}
\begin{center}
\includegraphics[scale=0.47]{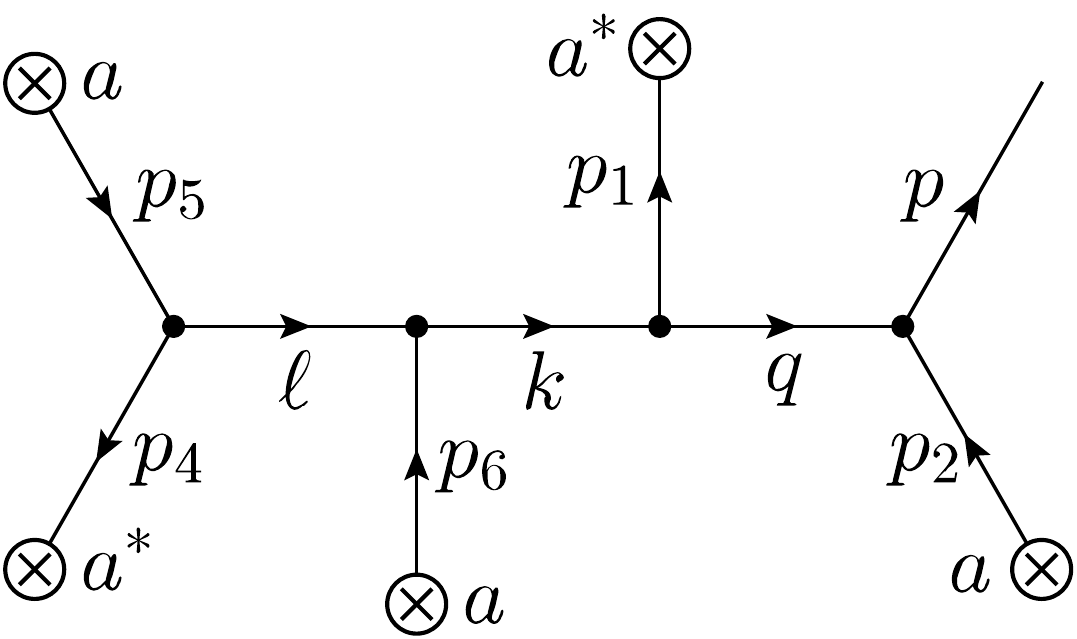}\\{\small (d)}
\end{center}
\end{minipage}
\begin{minipage}[b]{0.325\textwidth}
\begin{center}
\includegraphics[scale=0.47]{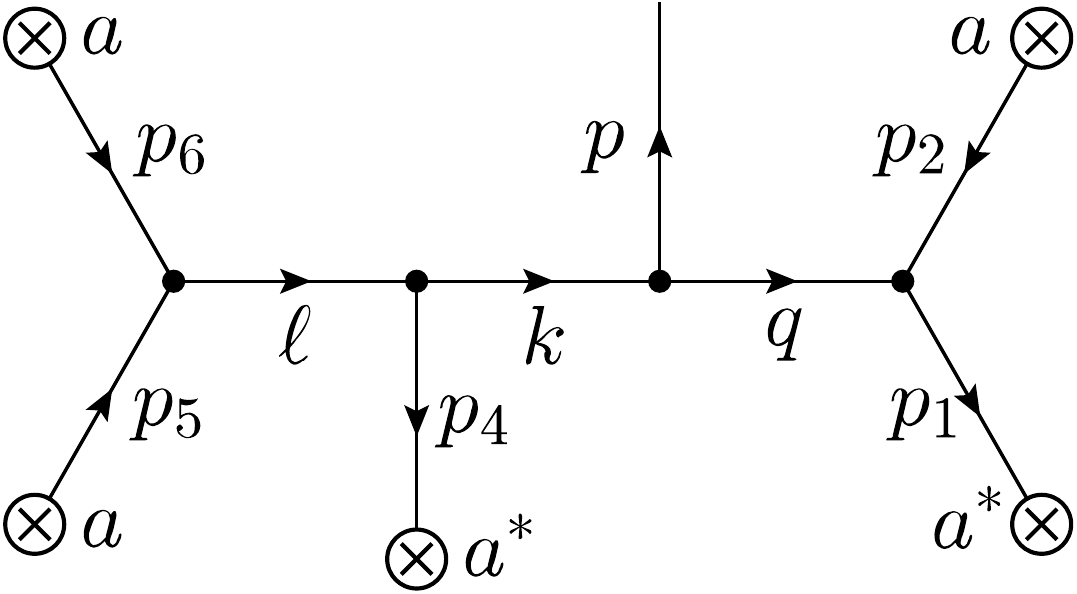}\\{\small (e)}
\end{center}
\end{minipage}
\begin{minipage}[b]{0.325\textwidth}
\begin{center}
\includegraphics[scale=0.47]{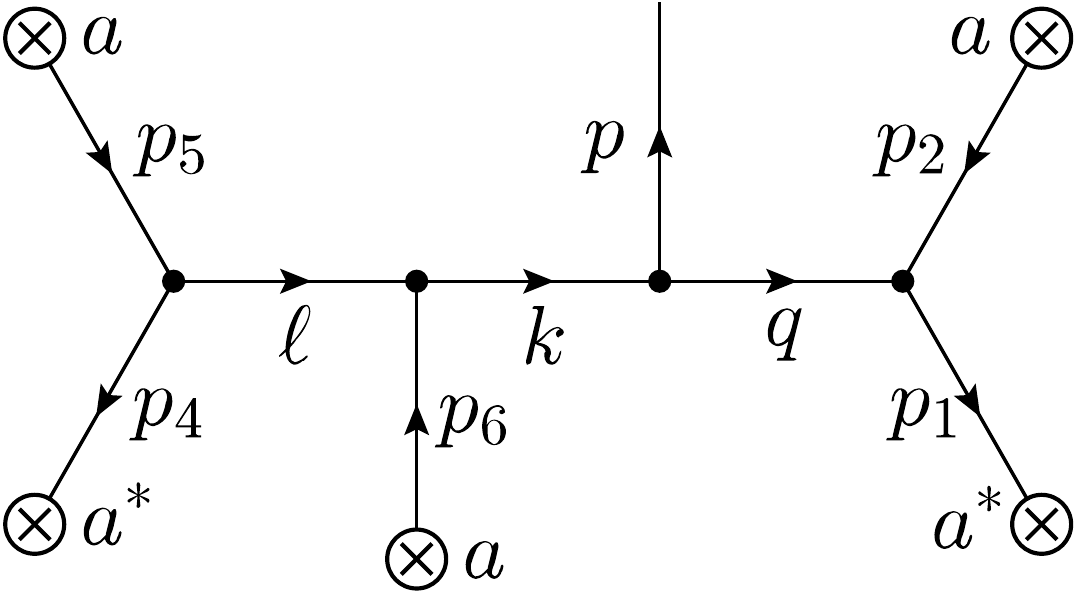}\\{\small (f)}
\end{center}
\end{minipage}
\caption{The $g^4$ contributions to $\delta a_{\bp}$ leading to the loss terms in the Boltzmann equation. From top left to bottom right, the first four diagrams arise from the first term of the current in \eqn{jg4} while the last two arise from the second one. The last term in \eqn{jg4} does not contribute to the Boltzmann equation.}
\label{fig:dapg4}
\end{figure}

Considering now the crossed term $a^*_{\bp'} \delta \dot{a}_{\bp}$ in \eqn{fpdot}, one needs to calculate $\delta a_{\bp}$ to order 
$g^4$. After straightforward iterations one finds that the current $J(y)$ to this order reads (in a compact notation where repeated coordinates are integrated over)
\beq
\label{jg4}
J_y = - \frac{g^4}{2}\, 
\phi_y \rmi\Delta_{yz} 
\phi_z \rmi\Delta_{zw} 
\phi_w \rmi\Delta_{wu} 
\phi^2_u
-\frac{g^4}{4}\,
\rmi\Delta_{yz} \phi_z 
\rmi\Delta_{zw} \phi_w^2
\rmi\Delta_{yu} \phi_u^2
-\frac{g^4}{8}\,
\phi_y \rmi\Delta_{yz}
\rmi\Delta_{zw} \phi_w^2
\rmi\Delta_{zu} \phi_u^2.
\eeq
In Fig.~\ref{fig:dapg4} we show the six diagrams contributing to $\delta a_{\bp}$. All corresponding expressions are very similar to \eqn{dapl2atemp} with the extra element of having two more propagators. We have
\begin{align}
\label{dapg4atemp}
\delta a_{\bp} = \,&
\frac{1}{h_{\bp}}\,\frac{g^4}{2}
\int \prod_{i} \frac{\dif^3 \bp_{i}}{h_{\bp_i}}\,
(2\pi)^3 \delta^{(3)}(\overline{\Delta\bp})\,
\frac{\rme^{\rmi (\overline{\Delta E} - \rmi \epsilon) x^0}}{\rmi(\overline{\Delta E} - \rmi \epsilon)}\,
a^*_{\bp_1} a_{\bp_2} a^*_{\bp_4} a_{\bp_5} a_{\bp_6}
\Delta_{\rm R}(p_5+p_6-p_4) 
\nn
&\big[\Delta_{\rm R}(p_5+p_6) 
\Delta_{\rm R} (p_5+p_6-p_4+p_2)
+\Delta_{\rm R}(p_5+p_6) 
\Delta_{\rm R} (p_5+p_6-p_4-p_1)
\nn
&+2\Delta_{\rm R}(p_5-p_4) 
\Delta_{\rm R} (p_5+p_6-p_4+p_2)
+2\Delta_{\rm R}(p_5-p_4) 
\Delta_{\rm R} (p_5+p_6-p_4 -p_1)
\nn
&+\Delta_{\rm R}(p_5+p_6) 
\Delta_{\rm R} (p_5+p_6-p_4-p)
+2\Delta_{\rm R}(p_5-p_4)
\Delta_{\rm R} (p_5+p_6-p_4-p)\big],
\end{align}
with $\overline{\Delta E}$ and $\overline{\Delta \bp}$ as in \eqn{dapl2atemp}.

At this point it is appropriate to say that only a propagator with argument the sum of three external momenta will have a real part leading to conservation of energy. In fact, we have already used this property when considering the propagators in \eqn{dapg2temp}; none of the two propagators there acquired a real part. Moreover, this is also the reason that no diagram coming from the last term of the current in \eqn{jg4} contributes to the Boltzmann equation; any propagator in such a diagram will have as an argument the sum of either two or four external momenta, as one can easily verify by simply drawing it. Thus, energy conservation will emerge out of \eqn{dapg4atemp}, as in \eqn{repropr}, from the propagator $\Delta_{\rm R}(p_5+p_6-p_4)$ and the only extra work we have to do is to carefully calculate the arguments of the remaining propagators in the square bracket in \eqn{dapg4atemp} after taking the ensemble average of $a^*_{\bp'} \delta \dot{a}_{\bp}$ and without worrying about their real parts. One has to always identify $\bp_4$ with $\bp_2$, while there is the possibility to choose $\bp_5 = \bp_1$, $\bp_6=\bp$ or $\bp_5 = \bp$, $\bp_6=\bp_1$. It is an easy exercise to verify that the sum of propagator products in the square bracket in \eqn{dapg4atemp} becomes
\beq
\big[{\textstyle \sum} \Delta_{\rm R}\Delta_{\rm R}\big]
\to  - 2 \left[ \frac{1}{s}+\frac{1}{t}+\frac{1}{u} \right]^2.
\eeq
Following now the exact same steps as in the case of the corresponding $\lambda^2$ term, and noticing in particular that the integrand is still invariant under $\bp_2 \leftrightarrow \bp_3$, and thus under $t \leftrightarrow u$, so that we can let $2 f_{\bp}f_{\bp_1}f_{\bp_2}$ $\to$ $f_{\bp}f_{\bp_1}f_{\bp_2} + f_{\bp}f_{\bp_1}f_{\bp_3}$, we arrive at the $g^4$ loss terms
\beq
\label{lossg4}
\dot{f}_{\bp}\big|_{g^4}^{B} = -\frac{1}{4 E_{\bp}} 
\int \widetilde{\dif \bp_1}\,
\widetilde{\dif \bp_2}\,
\widetilde{\dif \bp_3}\,
(2\pi)^4 \delta^{(4)}(\Delta p)\,
\left[\frac{g^2}{s}+\frac{g^2}{t}+\frac{g^2}{u} \right]^2
\big[f_{\bp}f_{\bp_1}f_{\bp_2} + f_{\bp}f_{\bp_1}f_{\bp_3}\big].
\eeq

Regarding the second $g^4$ gain term, we can draw four diagrams with the external lines $a_{\bp_1} a_{\bp_2} a^*_{\bp_4} a^*_{\bp_5} a_{\bp_6}$. Then, in analogy to the computation performed for diagram \ref{fig:dapl2}.b, it is not hard to convince ourselves that we get a contribution
\beq
\label{gaing4b}
\dot{f}_{\bp}\big|_{g^4}^{C} = \frac{1}{4 E_{\bp}} 
\int \widetilde{\dif \bp_1}\,
\widetilde{\dif \bp_2}\,
\widetilde{\dif \bp_3}\,
(2\pi)^4 \delta^{(4)}(\Delta p)\,
\left[\frac{g^2}{s}+\frac{g^2}{t}+\frac{g^2}{u} \right]^2
f_{\bp}f_{\bp_2}f_{\bp_3}.
\eeq

Now we put together all the $g^4$ contributions from Eqs.~\eqref{gaing4}, \eqref{lossg4} and \eqref{gaing4b} to arrive at the Boltzmann equation in the classical $\phi^3$ theory, that is
\beq
\hspace*{-0.15cm}
\label{allg4}
\dot{f}_{\bp}\big|_{g^4} \!=\! \frac{1}{4 E_{\bp}}\! 
\int \!\widetilde{\dif \bp_1}\,
\widetilde{\dif \bp_2}\,
\widetilde{\dif \bp_3}\,
(2\pi)^4 \delta^{(4)}(\Delta p)\!
\left[\frac{g^2}{s}+\frac{g^2}{t}+\frac{g^2}{u} \right]^2\!\!\!
\big[
 f_{\bp_2}f_{\bp_3} \big(f_{\bp_1} + f_{\bp}\big)   
- f_{\bp}f_{\bp_1}  \big(f_{\bp_2} + f_{\bp_3}\big) 
\big].
\eeq

\subsection{\label{subsec:lg2} The $\lambda g^2$ terms and the Boltzmann equation for the full scalar theory}

Finally, in order to complete the derivation of the Boltzmann equation in the full scalar theory, i.e.~with both cubic and quartic vertices, we need to compute the terms of order $\lambda g^2$.

The first gain term emerging from the product $\delta a^*_{\bp} \delta \dot{a}_{\bp}$ is rather easy to obtain since we already have the $\lambda$ and $g^2$ contributions to 
$\delta a_{\bp}$ as given in Eqs.~\eqref{dapltemp} and \eqref{dapg2temp} respectively. Compared to the corresponding calculation of the $\lambda^2$ and $g^4$ terms, the only difference in this mixed term is coming again from the propagators which after the ensemble average give
\beq
\Delta_{\rmR}(p_2+p_3) + 2 \Delta_{\rmR}(p_3-p_1)
- \Delta^*_{\rmR}(p'_2+p'_3) - 2 \Delta^*_{\rmR}(p'_3-p'_1)
\to 4\rmi\left[ \frac{1}{s} + \frac{1}{t} + \frac{1}{u} \right].
\eeq
In the above we have user for one more time our freedom to let $1/t \to 1/u$ due to the invariance of the integrand in the subsequent integrations. We finally find the gain term
\beq
\label{gainlg2}
\dot{f}_{\bp}\big|_{\lambda g^2}^{A} = \frac{1}{4 E_{\bp}} 
\int \widetilde{\dif \bp_1}\,
\widetilde{\dif \bp_2}\,
\widetilde{\dif \bp_3}\,
(2\pi)^4 \delta^{(4)}(\Delta p)\,
2 \lambda
\left[\frac{g^2}{s}+\frac{g^2}{t}+\frac{g^2}{u} \right]
f_{\bp_1}f_{\bp_2}f_{\bp_3}.
\eeq

\begin{figure}
\begin{minipage}[b]{0.325\textwidth}
\begin{center}
\includegraphics[scale=0.45]{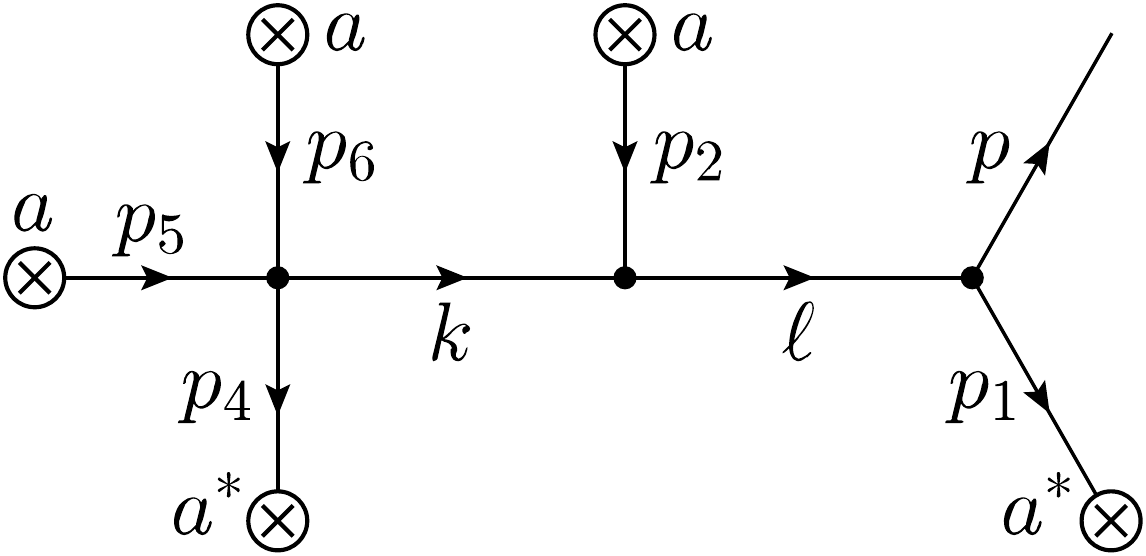}\\{\small (a)}
\end{center}
\end{minipage}
\begin{minipage}[b]{0.325\textwidth}
\begin{center}
\includegraphics[scale=0.45]{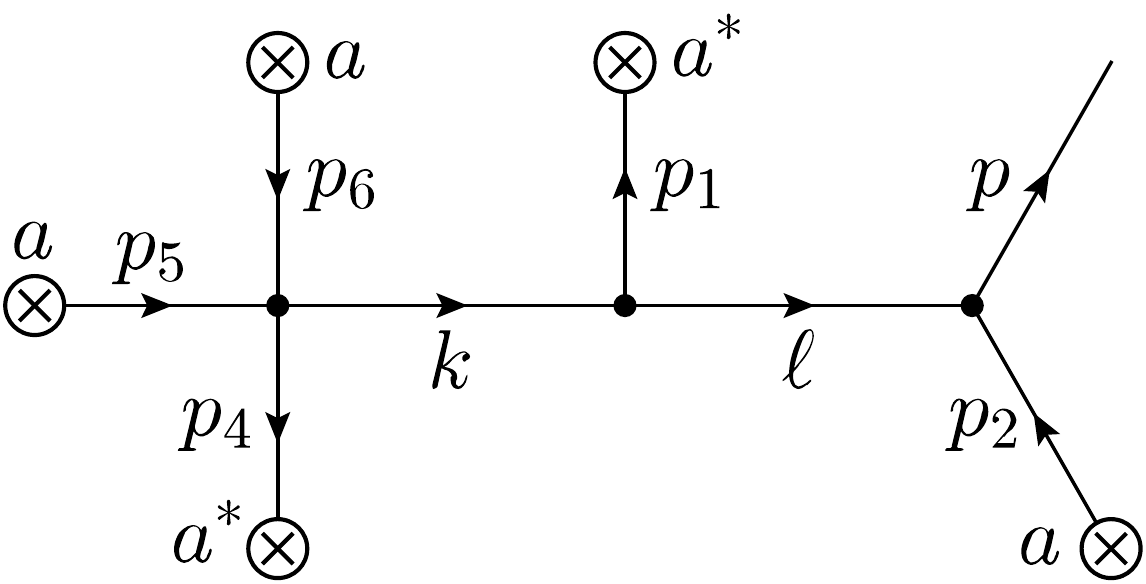}\\{\small (b)}
\end{center}
\end{minipage}
\begin{minipage}[b]{0.325\textwidth}
\begin{center}
\includegraphics[scale=0.45]{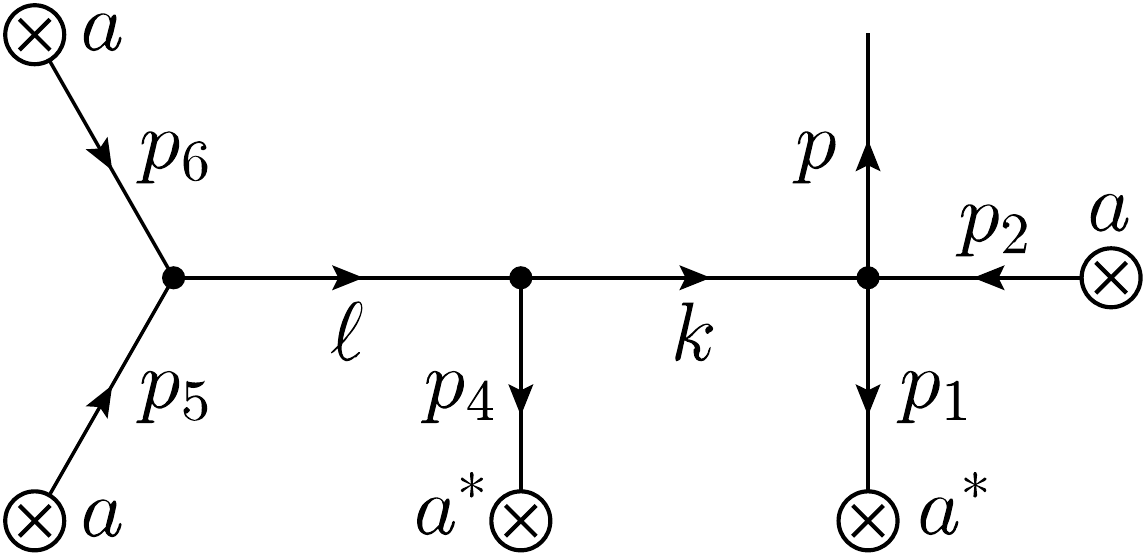}\\{\small (c)}
\end{center}
\end{minipage}
\begin{minipage}[b]{0.35\textwidth}
\vspace{0.5cm}
\begin{center}
\includegraphics[scale=0.45]{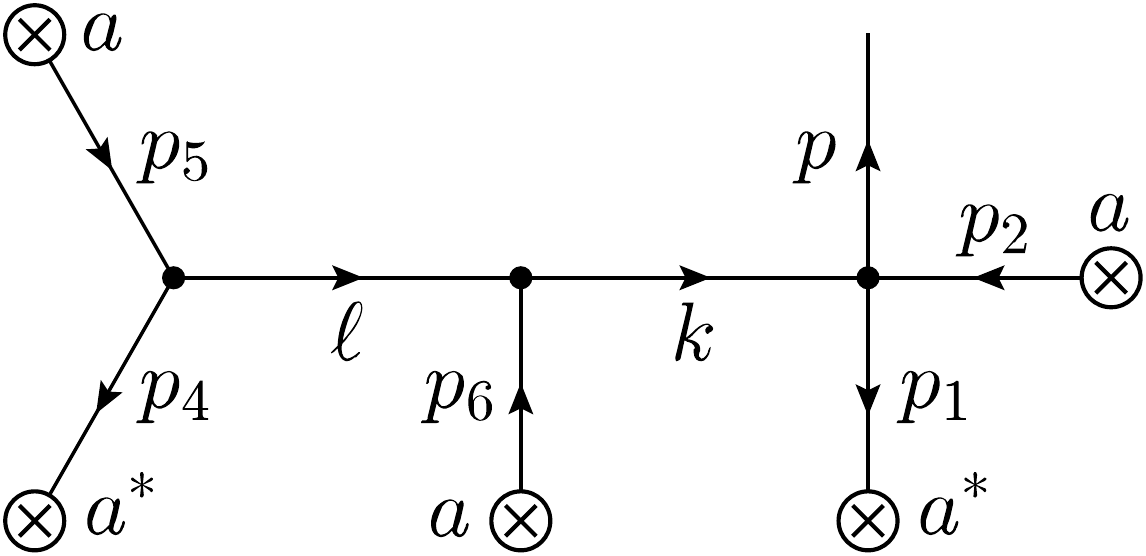}\\{\small (d)}
\end{center}
\end{minipage}
\begin{minipage}[b]{0.35\textwidth}
\begin{center}
\includegraphics[scale=0.45]{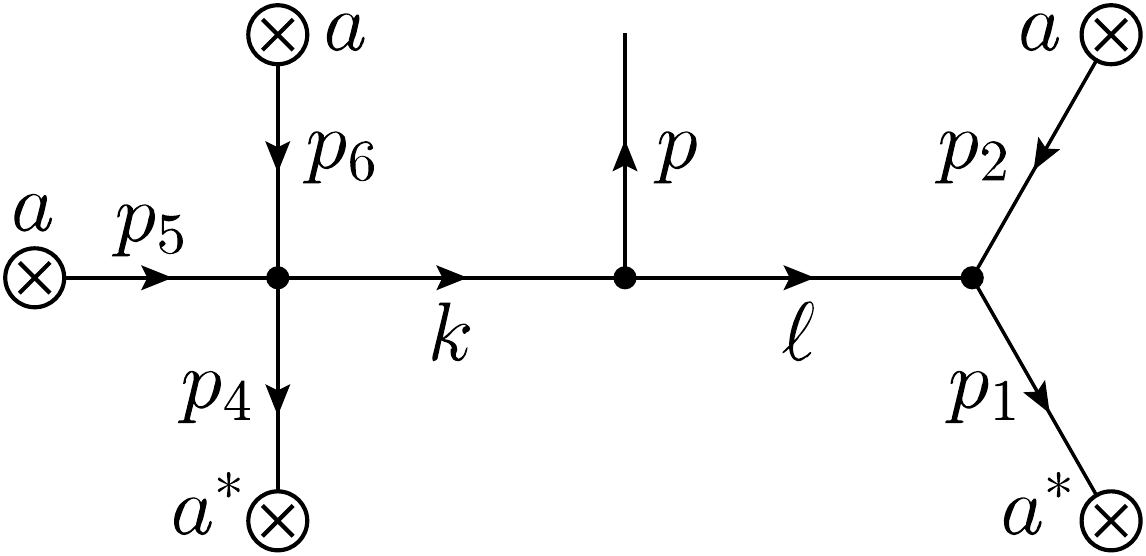}\\{\small (e)}
\end{center}
\end{minipage}
\caption{The $\lambda g^2$ contributions to $\delta a_{\bp}$ leading to the loss terms in the Boltzmann equation. From top left to bottom right, the first two diagrams arise from the first term of the current in \eqn{jlg2}, the next two from the second one  while the last one comes from the third one. The last two terms in \eqn{jlg2} do not contribute to the Boltzmann equation.}
\label{fig:daplg2}
\end{figure}

Regarding the crossed term $a^*_{\bp'} \delta \dot{a}_{\bp}$ term in \eqn{fpdot}, we have to compute
$\delta a_{\bp}$ to order $\lambda g^2$. After straightforward iterations we finds that the current $J(y)$ to this order is given by
\begin{align}
\label{jlg2}
J_y = &- \frac{\lambda g^2}{6}\, 
\phi_y \rmi\Delta_{yz} 
\phi_z \rmi\Delta_{zw} 
\phi_w^3
-\frac{\lambda g^2}{4}\, 
\phi_y^2 \rmi\Delta_{yz} 
\phi_z \rmi\Delta_{zw} 
\phi_w^2
-\frac{\lambda g^2}{12}\, 
\rmi\Delta_{yz} 
\phi_z^2 \rmi\Delta_{yw} 
\phi_w^3
\nn
&-\frac{\lambda g^2}{4}\, 
\phi_y \rmi\Delta_{yz} 
\phi_z^2 \rmi\Delta_{zw} 
\phi_w^2
-\frac{\lambda g^2}{8}\, 
\phi_y \rmi\Delta_{yz} 
\phi_z^2 \rmi\Delta_{yw} 
\phi_w^2.
\end{align}
In Fig.~\ref{fig:daplg2} we present the five diagrams contributing to $\delta a_{\bp}$ to order $\lambda g^2$. All corresponding expressions have similar structure to that of Eqs.~\eqref{dapl2atemp} and \eqref{dapg4atemp}, more precisely we have
\begin{align}
\label{daplg2atemp}
\delta a_{\bp} = \,& 
\frac{\rmi}{h_{\bp}}\,\frac{\lambda g^2}{2}
\int \prod_{i} \frac{\dif^3 \bp_{i}}{h_{\bp_i}}\,
(2\pi)^3 \delta^{(3)}(\overline{\Delta\bp})\,
\frac{\rme^{\rmi (\overline{\Delta E} - \rmi \epsilon) x^0}}{\rmi(\overline{\Delta E} - \rmi \epsilon)}\,
a^*_{\bp_1} a_{\bp_2} a^*_{\bp_4} a_{\bp_5} a_{\bp_6}
\Delta_{\rm R}(p_5+p_6-p_4) 
\nn
&\big[\Delta_{\rm R}(p_5+p_6-p_4+p_2) 
+\Delta_{\rm R}(p_5+p_6-p_4-p_1)
+\Delta_{\rm R}(p_5+p_6)
+2\Delta_{\rm R}(p_5-p_4) 
\nn*[0.2cm]
&+\Delta_{\rm R}(p_5+p_6-p_4-p) \big],
\end{align}
with $\overline{\Delta E}$ and $\overline{\Delta \bp}$ as in \eqn{dapl2atemp}.

Energy conservation will come from the propagator $\Delta_{\rm R}(p_5+p_6-p_4)$ as in the respective $\lambda^2$ and $g^4$ terms. Taking the ensemble average in the product $a^*_{\bp'} \delta \dot{a}_{\bp}$ we will identify $\bp_4$ with $\bp_2$, while there is the possibility to choose $\bp_5 = \bp_1$, $\bp_6=\bp$ or $\bp_5 = \bp$, $\bp_6=\bp_1$. Then the propagator bracket in \eqn{daplg2atemp} becomes
\beq
\big[ {\textstyle \sum} \Delta_{\rmR} \big] 
\to
4 \rmi \left[\frac{1}{s} + \frac{1}{t} + \frac{1}{u} \right].
\eeq
Now we copy the same steps as in the case of the corresponding $\lambda^2$ and $g^4$ terms, to arrive at the $\lambda g^2$ loss terms
\beq
\label{losslg2}
\dot{f}_{\bp}\big|_{\lambda g^2}^{B} = -\frac{1}{4 E_{\bp}} 
\int \widetilde{\dif \bp_1}\,
\widetilde{\dif \bp_2}\,
\widetilde{\dif \bp_3}\,
(2\pi)^4 \delta^{(4)}(\Delta p)\,
2 \lambda
\left[\frac{g^2}{s}+\frac{g^2}{t}+\frac{g^2}{u} \right]
\big[f_{\bp}f_{\bp_1}f_{\bp_2} + f_{\bp}f_{\bp_1}f_{\bp_3}\big].
\eeq

Concerning the second $\lambda g^2$ gain term, we can draw four diagrams with the external lines $a_{\bp_1} a_{\bp_2} a^*_{\bp_4} a^*_{\bp_5} a_{\bp_6}$. Then, in analogy to the previous respective computations we get the expected contribution
\beq
\label{gainlg2b}
\dot{f}_{\bp}\big|_{\lambda g^2}^{C} = \frac{1}{4 E_{\bp}} 
\int \widetilde{\dif \bp_1}\,
\widetilde{\dif \bp_2}\,
\widetilde{\dif \bp_3}\,
(2\pi)^4 \delta^{(4)}(\Delta p)\,
2 \lambda
\left[\frac{g^2}{s}+\frac{g^2}{t}+\frac{g^2}{u} \right]
f_{\bp}f_{\bp_2}f_{\bp_3}.
\eeq

It is trivial to add Eqs.~\eqref{gainlg2}, \eqref{losslg2} and \eqn{gainlg2b} to get the total $\lambda g^2$ contribution. By furthermore adding the total $\lambda^2$ and $g^4$ expressions given in Eqs.~\eqref{alll2} and \eqref{allg4}, we come to the Boltzmann equation for the full scalar theory
\beq
\label{fullbe}
\dot{f}_{\bp} = \frac{1}{4 E_{\bp}} 
\int \widetilde{\dif \bp_1}\,
\widetilde{\dif \bp_2}\,
\widetilde{\dif \bp_3}\,
(2\pi)^4 \delta^{(4)}(\Delta p)\,
|\mcal{M}|_{\phi}^2
\big[ f_{\bp_2}f_{\bp_3} \big(f_{\bp_1} + f_{\bp}\big)   
- f_{\bp}f_{\bp_1}  \big(f_{\bp_2} + f_{\bp_3}\big) 
\big],
\eeq
where we have defined the scattering amplitude squared of the full scalar theory
\beq
\label{sasf3f4}
|\mcal{M}|_{\phi}^2 = \left[\lambda + 
\frac{g^2}{s} + \frac{g^2}{t} + \frac{g^2}{u} \right]^2.
\eeq
Here we would like to stress that the specific combination of the occupation numbers in \eqn{fullbe} and the scattering amplitude squared of the scalar theory have emerged as a result of our calculation. Let us also notice that a factor 1/2 in front of the integral in \eqn{fullbe} is a symmetry factor due to the fact that particles 2 and 3, whose momenta are integrated over, are identical.

Furthermore, notice that the explicit form of 
$|\mcal{M}|_{\phi}^2$ as given in \eqn{sasf3f4} was derived in detail in the context of this scalar field theory. In the Yang-Mills case, which follows in the next section, we shall not derive the  respective amplitude squared $|\mcal{M}|_{\rm YM}^2$, since this is a standard, albeit not trivial, textbook calculation. However, we shall of course show that $|\mcal{M}|_{\rm YM}^2$ emerges in all terms in the Boltzmann equation and this is sufficient for our proof. Thus, it is useful to reflect back and see how we arrived at $|\mcal{M}|_{\phi}^2$ in this section. This is straightforward for the diagonal gain term; combining Eqs.~\eqref{dapltemp} and \eqref{dapg2temp} we see that 
$\mcal{M}_{\phi}(p_2 p_3; p p_1)$ appears in the integrand in the DA. Similarly $\mcal{M}^*_{\phi}(p'_2 p'_3 ; p p'_1)$ appears in the CCA and after squaring and performing the ensemble average we arrive at $|\mcal{M}(p_2 p_3 ; p p_1)|_{\phi}^2$. Regarding the crossed term it is enough to look, for example, in the loss terms and a first discussion  has already appeared below \eqn{lossl2} in the $\lambda \phi^4$ case. Putting together Eqs.~\eqref{dapl2atemp}, \eqref{dapg4atemp} and \eqref{daplg2atemp} we see that $\mcal{M}_{\phi}(p_2 k ; p p_1) \mcal{M}^{*}_{\phi}(p_4 k ; p_5 p_6)$ appears in the DA. After multiplying with the CCA taking the ensemble average and using the fact that $k$ is put on-shell according to \eqn{repropr} we arrive again at $|\mcal{M}(p_2 p_3 ; p p_1)|_{\phi}^2$ (cf.~the renaming of the momentum $k$ below \eqn{repropr}).        

%%%%%%%%%%%%%%%%%%%%%%%%%%%%%%%%%%%%%%%%%%%%%%%%%
%% 					QCD
%%%%%%%%%%%%%%%%%%%%%%%%%%%%%%%%%%%%%%%%%%%%%%%

\section{\label{sec:ym}Yang-Mills theory}

Now we would like to extend our analysis to the Yang-Mills theory in $D=4$ dimensions. Even though we will keep the number of colors $N_c$ arbitrary, we shall refer to the gauge bosons as gluons. The topology of the diagrams is the same as that in the full scalar theory studied in Sect.~\ref{sec:scalar} and the extra complications come only from the color and spin structure of the diagrams. The Yang-Mills action in an axial gauge reads
 \beq
 \label{sym}
 S_{\rm YM} = \int \dif^4 x \, \mcal{L}_{\rm YM} = 
 \int \dif^4 x  
 \left[ 
 -\frac{1}{4}\, F_{\mu\nu}^a F_a^{\mu\nu}
 -\frac{1}{2 \xi}\, \big(n^{\mu} A_{\mu}^{a}\big)^2
 \right],
 \eeq 
with the field strength
 \beq
 \label{fmunua}
 F_{\mu\nu}^a = \del_{\mu} A^a_{\nu}
 - \del_{\nu} A^a_{\mu}
 +g f^{abc} A^b_{\mu} A^c_{\nu}
 \eeq
and where $f^{abc}$ are the familiar structure constants of the $SU(N_c)$ group. In general, $n^{\mu}$ and $\xi$ are arbitrary in \eqn{sym}, but for our convenience we shall consider the light-cone gauge defined by the conditions $n^{\mu}n_{\mu} = 0$ and $\xi \to 0$. Introducing the polarization vectors 
$\varepsilon^{\lambda}_{\mu}(\bp)$ for the two transverse (and physical) gluon polarizations, and which satisfy $p \cdot \varepsilon^{\lambda}(\bp)= n \cdot\varepsilon^{\lambda}(\bp)=0$, we can expand the gauge field as
 \beq
 \label{amua}
 A_{\mu}^a(x)=
 \int \frac{\dif^3 \bp}{h_{\bp}}\,
 \left[
 a_{\bp}^{\lambda a}\,
 \varepsilon^{\lambda}_{\mu}(\bp)\,
 \rme^{-\rmi p \cdot x}
 +
 a_{\bp}^{\lambda a *}\,
 \varepsilon^{\lambda *}_{\mu}(\bp)\,
 \rme^{\rmi p \cdot x}
 \right].  
 \eeq 
Assuming $a_{\bp}^{\lambda a}$ is slowly varying and using the orthogonality property of the polarization vectors, i.e.~$\varepsilon_{\lambda}(\bp) \cdot\varepsilon_{\lambda'}^*(\bp) = -\delta_{\lambda \lambda'}$ one can invert the above to find
 \beq
 \label{apla}
 a_{\bp}^{\lambda a} = 
 -\frac{\rmi}{h_{\bp}}
 \int \dif^3 \bx\, \rme^{\rmi p \cdot x}
 \varepsilon^{\lambda *}_{\mu}(\bp)
 [\dot{A}^{\mu a}(x) - \rmi E_{\bp} A^{\mu a}(x)].
 \eeq
Apart from the consideration of a homogeneous medium, we will also assume that the occupation numbers are independent of color and spin, that is, 
 \beq
 \label{astaraym}
 \big\langle
 \big(a^{\lambda' a'}_{\bp'}\big)^*
 a^{\lambda a}_{\bp}
 \big\rangle
 = \delta^{\lambda\lambda'}
 \delta^{aa'}
 \delta^{(3)}_{\bp\bp'}\, f_{\bp}.
 \eeq
In order to follow the classical evolution of the system, we need the corresponding equations of motion which read
 \beq
 \label{boxa}
 \bigg(g^{\mu\nu}\Box - 
 \del^{\mu} \del^{\nu} 
 -\frac{1}{\xi}\,
 n^{\mu}n^{\nu}\bigg)
 A^a_{\nu}
 = J^{\mu a}(x)
 \eeq 
with a current having quadratic and cubic terms in the gauge fields 
 \beq
 J_\mu^a= - 
 g f^{abc}
 \Big[\big(\del^\nu A^b_\nu\big) A^c_\mu 
 + 2 A^b_\nu\, \del^\nu A^c_\mu 
 - A^b_\nu\, \del_\mu A^{\nu c} \Big]
 - g^2 f^{abe}f^{cde} 
 A_{\nu}^b A^{\nu c} A^d_\mu.  
 \eeq
Now we expand the full interacting field according to $A_\mu^a = A_\mu^{(0)a}+ \delta A_\mu^a$, with $A_\mu^{(0)a}$ a free field and $\delta A_\mu^a$ the piece induced by the interactions and given by
 \beq
 \label{deltaamua}
 \delta A^a_{\mu}(x) =
 -\int \dif^4 y\, 
 \rmi G_{\mu\nu}(x-y) 
 J^{\nu a}(y),
 \eeq  
where we have already used the fact that the propagator is diagonal in color. It is taken to be the retarded one, and in momentum space in the light-cone gauge it reads
 \beq
 \label{grprop}
 G_{\rm R}^{\mu\nu}(k) = 
 \frac{\rmi}{k^2 + \rmi \epsilon k^0} 
 \,\left(-g^{\mu\nu} + 
 \frac{n^\mu k^{\nu} + n^{\nu}k^{\mu}}{n\cdot k}  
 \right),
 \eeq
where $\epsilon \to 0^+$, while the prescription for the axial pole is irrelevant for our purposes\footnote{It cannot give rise to real parts leading to energy conservation as, for example, in \eqn{repropr}.}. Now expanding $a_{\bp}^{\lambda a} = a_{\bp}^{(0)\lambda a} + \delta a_{\bp}^{\lambda a}$ one finds that the change in the field coefficients is given by\footnote{To that aim, one has to make use of \eqn{proppol} which appears below in Sect.~\ref{sec:crossedym}.}
 \beq
 \label{dapla}
 \delta a_{\bp}^{\lambda a}  
 = - \frac{\rmi}{h_{\bp}} 
 \int \dif^4 y\, 
 \rme^{\rmi p \cdot y}\,
 \Theta(x^0-y^0)\,
 \varepsilon_{\mu}^{\lambda *}(\bp) 
 J^{\mu a}(y). 
 \eeq
Finally the occupation numbers evolve in time according to
 \beq
 \label{fpdotym}
 \delta^{aa'}
 \delta^{\lambda\lambda'} 
 \delta^{(3)}_{\bp\bp'}
 \,\dot{f}_{\bp}
 = 2 \mathrm{Re} 
 \big[\big\langle 
 a^{\lambda' a' *}_{\bp'} 
 \delta \dot{a}^{\lambda a}_{\bp}
 \big\rangle \big]
 + 2 \mathrm{Re} 
 \big[ \big\langle
 \delta a^{\lambda' a'*}_{\bp'} 
 \delta \dot{a}^{\lambda a}_{\bp} 
 \big\rangle\big],
 \eeq
where we have already dropped the superscript $(0)$ in the field coefficients.

\subsection{\label{sec:diagym} The Feynman rules for the classical Yang-Mills theory and the diagonal, gain, term}

Before proceeding to calculate the diagonal contribution to \eqn{fpdotym} let us establish the Feynman rules for the calculation of $\delta a^{\lambda a}_{\bp}$. Most of the rules remain the same as the corresponding ones in the scalar theory, while we have the modifications listed below.  
\begin{list}{\small $\Box$}{\setlength{\itemsep}{0cm} \setlength{\itemindent}{0cm}}
\item Assign a factor $V^{abc}_{\mu\nu\rho}(p_1,p_2,p_3)$ for each cubic vertex and a factor $V^{abcd}_{\mu\nu\rho\sigma}$ for each quartic one where
 \begin{align}
 V^{abc}_{\mu\nu\rho}(p_1,p_2,p_3)
 = g f^{abc} 
 \big[ 
 g_{\mu\nu}(p_1 -p_2)_\rho +
 g_{\nu\rho} (p_2 -p_3)_\mu  + 
 g_{\rho\mu} (p_3 -p_1)_\nu
 \big],
 \end{align}
 \begin{align}
 V^{abcd}_{\mu\nu\rho\sigma} =
 -\rmi g^2
 \big[
 &f^{abe} f^{cde}
 (g_{\mu\rho} g_{\nu\sigma}
 -g_{\mu\sigma}g_{\nu\rho})
 +f^{ace} f^{bde}
 (g_{\mu\nu} g_{\rho\sigma}
 -g_{\mu\sigma}g_{\nu\rho})
 \nn
 +&f^{ade} f^{bce}
 (g_{\mu\nu} g_{\rho\sigma}
 -g_{\mu\rho}g_{\nu\sigma})
 \big] .
 \end{align}
\item Use the retarded propagator $G^{\mu\nu}_{\mathrm R}(k)$ given in \eqn{grprop} for each internal line with the four-momentum $k$ flowing towards the measured occupation factor. Equivalently, one can use the advanced propagator $G^{\mu\nu}_{\mathrm A}(k) = G^{\mu\nu}_{\mathrm R}(-k)$ if the four-momentum $k$ is taken to flow away from the measured occupation factor.
\item Integrate according to $\displaystyle{\int \frac{\dif^3 \bp}{h_{\bp}}}\, a_{\bp}^{\lambda a *}\,
 \varepsilon^{\lambda *}_{\mu}(\bp)$ or
$\displaystyle{\int \frac{\dif^3 \bp}{h_{\bp}}}\, a_{\bp}^{\lambda a}\,
 \varepsilon^{\lambda}_{\mu}(\bp)$ for each external line.
 \item Multiply by $\varepsilon^{\lambda *}_{\mu}(\bp)$ for the measured momentum.  
\end{list}
Again, we stress that these rules represent in a convenient way the perturbative solution to the classical equation of motion. 

\begin{figure}
\begin{minipage}[b]{0.22\textwidth}
\begin{center}
\includegraphics[scale=0.52]{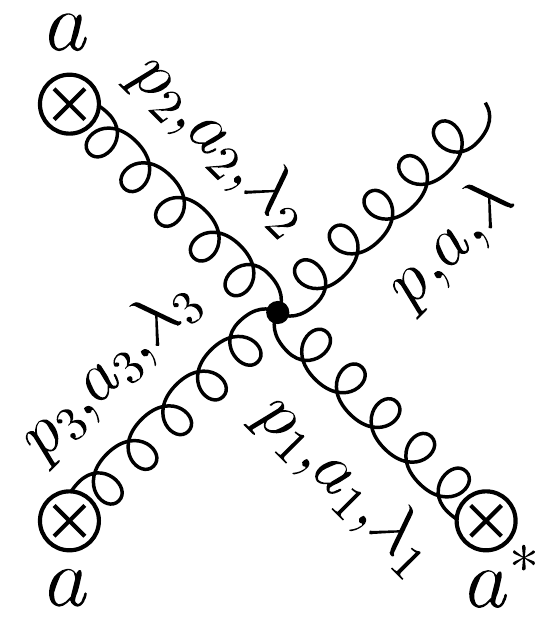}\\*[0.8cm]{\small (a)}
\end{center}
\end{minipage}
\begin{minipage}[b]{0.32\textwidth}
\begin{center}
\includegraphics[scale=0.52]{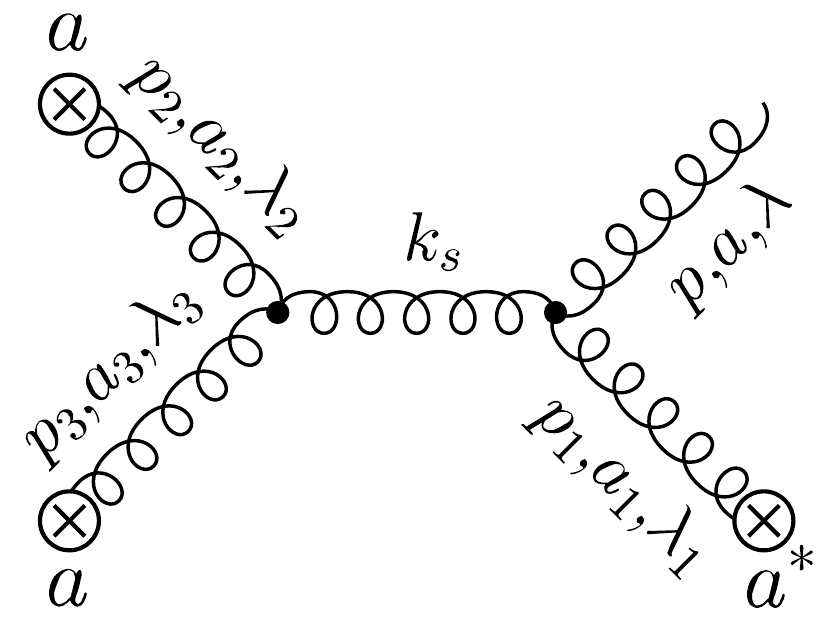}\\*[0.8cm]{\small (b)}
\end{center}
\end{minipage}
\begin{minipage}[b]{0.22\textwidth}
\begin{center}
\includegraphics[scale=0.52]{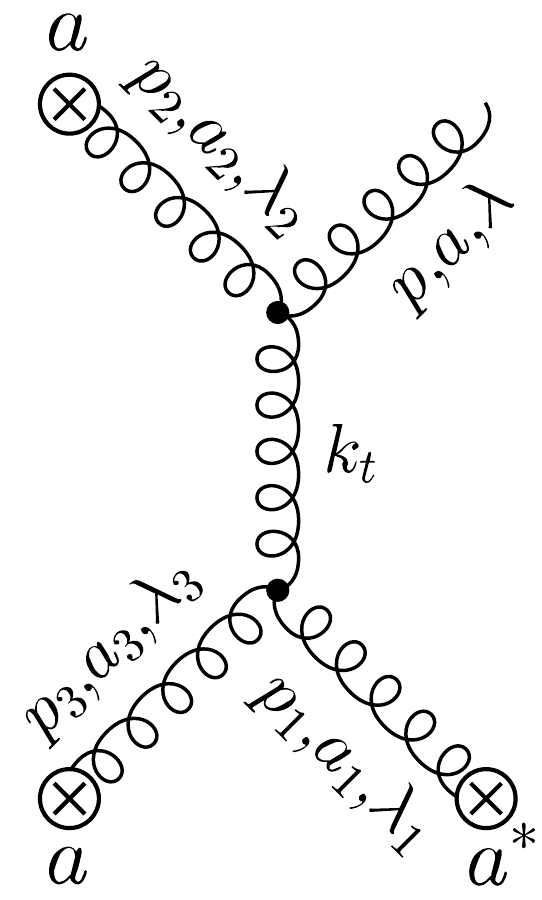}\\{\small (c)}
\end{center}
\end{minipage}
\begin{minipage}[b]{0.22\textwidth}
\begin{center}
\includegraphics[scale=0.52]{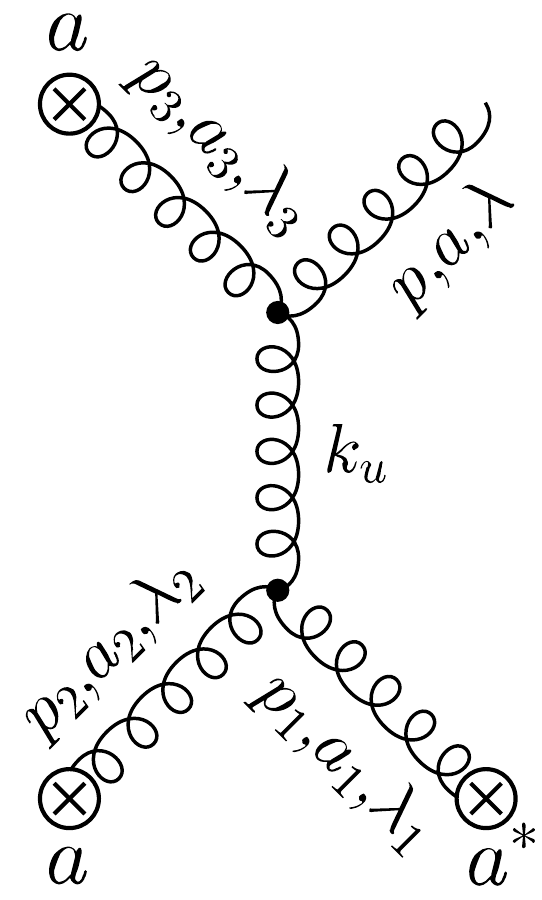}\\{\small (d)}
\end{center}
\end{minipage}
\caption{The $g^2$ contribution to $\delta a_{\bp}$, cf.~\eqn{dapdiag}. Momenta $p_2$ and $p_3$ flow inwards, while momenta $p$ and $p_1$ flow outwards in all four diagrams. The momentum of the exchanged gluon in diagrams (b), (c) and (d) flows towards the measured occupation number $f_{\bp}$.}
\label{fig:dapym}
\end{figure}

Thus, in view of calculating the diagonal term, let us begin with the $g^2$ contribution to $\delta a_{\bp}^{\lambda a}$ due to both cubic and quartic interactions. Recall that in the DA we need to keep only the $a^*aa$ term to satisfy energy conservation and, with the notation we used in the scalar theory, we have
 \beq
 \label{dapdiag}
 \delta a_{\bp}^{\lambda a} = 
 \frac{1}{2 h_{\bp}}
 \int \prod_{i} 
 \frac{\dif^3 \bp_i}{h_{\bp_i}}\,
 (2\pi)^3 \delta^{(3)}(\Delta \bp)
 \frac{\rme^{\rmi (\Delta E- \rmi 
 \epsilon)x^0}}{\rmi (\Delta E- \rmi 
 \epsilon)}
 a_{\bp_1}^{\lambda_1 a_1*}
 a_{\bp_2}^{\lambda_2 a_2}
 a_{\bp_3}^{\lambda_3 a_3}
 \mcal{M}_{\lambda\lambda_1 a a_1}^{\lambda_2\lambda_3 a_2 a_3}(p_2 p_3 ; p p_1),
 \eeq 
where a summation over repeated color and spin indices is understood. In \eqn{dapdiag} we have defined the total amplitude for $2\to2$ scattering
$\mcal{M}_{\lambda\lambda_1 a a_1}^{\lambda_2\lambda_3 a_2 a_3}(p_2 p_3 ;  p p_1)$. This is the sum of contributions  involving three-gluon and four-gluon interactions which are given by
 \begin{align}
 \label{m3g}
 \mcal{M}_{\lambda\lambda_1 a a_1}^{\lambda_2\lambda_3 a_2 a_3}(p_2 p_3; p p_1)\big|_{3g} =
 \,\,&\varepsilon^{\lambda*}_{\mu}(\bp)
 \varepsilon^{\lambda_1*}_{\mu_1}(\bp_1)
 \varepsilon^{\lambda_2}_{\mu_2}(\bp_2)
 \varepsilon^{\lambda_3}_{\mu_3}(\bp_3)
 \nn
 \times\, \big[
 &\,V_{a_2 a_3 b}^{\mu_2\mu_3\nu}(p_2,p_3,-k_s)\,
 G^{\rm R}_{\nu\rho}(k_s)\,
 V_{a a_1 b}^{\mu \mu_1 \rho}(-p,-p_1,k_s)
 \nn +
 &\,V_{a_3 a_1 b}^{\mu_3\mu_1\nu}(p_3,-p_1,-k_t)\,
 G^{\rm R}_{\nu\rho}(k_t)\,
 V_{a_2 a b}^{\mu_2 \mu \rho}(p_2,-p,k_t)
 \nn + 
 &\,V_{a_2 a_1 b}^{\mu_2\mu_1\nu}(p_2,-p_1,-k_u)\,
 G^{\rm R}_{\nu\rho}(k_u)\,
 V_{a a_3 b}^{\mu \mu_3 \rho}(p_3,-p,k_u)
 \big],
 \end{align}
 \beq
 \label{m4g}
 \mcal{M}_{\lambda\lambda_1 a a_1}^{\lambda_2\lambda_3 a_2 a_3}(p_2 p_3; p p_1)\big|_{4g}=
 \varepsilon^{\lambda*}_{\mu}(\bp)
 \varepsilon^{\lambda_1*}_{\mu_1}(\bp_1)
 \varepsilon^{\lambda_2}_{\mu_2}(\bp_2)
 \varepsilon^{\lambda_3}_{\mu_3}(\bp_3)\,
 V^{\mu\mu_1\mu_2\mu_3}_{aa_1a_2a_3},
 \eeq
where in \eqn{m3g} we have defined the four-momenta $k_s=p_2+p_3$,  $k_t = p_3-p_1$ and $k_u=p_2-p_1$ and the three terms in the square bracket clearly correspond to the $s$, $t$ and $u$ diagrams. The four diagrams contributing to \eqn{dapdiag} are shown 
in Fig.~\ref{fig:dapym}.

From the (DA) in \eqn{dapdiag} we easily build its time derivative and the CCA. Denoting all momenta in the CCA with a prime, we form the diagonal term in \eqn{fpdotym} and we assume the ensemble average (cf.~the analogous \eqn{fact} in the scalar theory)
\begin{align}
\label{factym}
\hspace*{-1.02cm}\Big\langle a^{\lambda_1 a_1*}_{\bp_1} a^{\lambda_2 a_2}_{\bp_2} 
a^{\lambda_3 a_3}_{\bp_3} 
a^{\lambda'_1 a'_1}_{\bp'_1} 
a^{\lambda'_2 a_2*}_{\bp'_2} 
a^{\lambda'_3 a'_3*}_{\bp'_3}
& \Big\rangle
\to
2 \Big\langle a^{\lambda_1 a_1*}_{\bp_1} a^{\lambda'_1 a'_1}_{\bp'_1} \Big\rangle 
\Big\langle a^{\lambda_2 a_2*}_{\bp_2} a^{\lambda'_2 a'_2}_{\bp'_2} \Big\rangle
\Big\langle a^{\lambda_3 a_3*}_{\bp_3} a^{\lambda'_3 a'_3}_{\bp'_3} \Big\rangle  
\nn
=&\, 
2\delta^{\lambda_1\lambda'_1}
\delta^{\lambda_2\lambda'_2}
\delta^{\lambda_3\lambda'_3}
\delta^{a_1 a'_1}
\delta^{a_2 a'_2}
\delta^{a_3 a'_3}
\delta^{(3)}_{\bp_1\bp'_1}
\delta^{(3)}_{\bp_2\bp'_2} 
\delta^{(3)}_{\bp_3\bp'_3}
f_{\bp_1}f_{\bp_2}f_{\bp_3}. 
\end{align} 
The factor of 2 in the above comes about because there are two possible contractions when performing the ensemble average, 
\{$1'=1$, $2'=2$, $3'=3$\} and \{$1'=1$, $2'=3$, $3'=2$\}, which give the same result since one is integrating over the external momenta and summing over the color and polarization indices. Then we encounter the product of the amplitudes in the DA and the CCA which becomes
 \beq
 \label{mmstar}
 \mcal{M}_{\lambda\lambda_1 a a_1}^{\lambda_2\lambda_3 a_2 a_3}
 (p_2 p_3 ;  p p_1)
 \Big[\mcal{M}_{\lambda'\lambda_1 a' a_1}^{\lambda_2\lambda_3 
 a_2  a_3}(p_2 p_3 ;  p p_1)\Big]^* =
 \frac{\delta^{\lambda\lambda'}}{2}\,
 \frac{\delta^{aa'}}{N_c^2-1}\,
 |\mcal{M}|^2_{\rm YM}.
 \eeq
In the above, $|\mcal{M}|^2_{\rm YM}$ is the scattering amplitude squared, summed over all initial and final colors and polarizations, at order $g^4$ in the Yang-Mills theory and it reads\footnote{For example, see Chapter 81 in \cite{Srednicki:2007qs} and in particular Eq.~(81.44).}
 \beq
 \label{m2ym}
 |\mcal{M}|^2_{\rm YM} = 
 16 N_c^2 (N_c^2-1) g^4
 \left[\frac{s^2-tu}{s^2} + 
 \frac{t^2-us}{t^2} + 
 \frac{u^2-st}{u^2} \right]. 
 \eeq
The remaining parts of the calculation are identical to those in the scalar theory (cf., for example, Sect.~\ref{subsec:l2}) and putting everything together we find the first gain term in the Boltzmann equation in the Yang-Mills theory, that is    
 \beq\label{beyma} 
 \dot f_{\bm p}\big|^A =  
 \frac{1}{4 E_{\bp}} 
 \int \widetilde{\dif \bp_1}\,
 \widetilde{\dif \bp_2}\,
 \widetilde{\dif \bp_3}\,
 (2\pi)^4 \delta^{(4)}(\Delta p)\,
 \frac{|\mcal{M}|^2_{\rm YM}}{2(N_c^2-1)} 
 f_{\bp_1}f_{\bp_2}f_{\bp_3}.
 \eeq
 
\subsection{\label{sec:crossedym}Loss and gain from the crossed term}

\begin{figure}
\begin{minipage}[b]{0.49\textwidth}
\begin{center}
\includegraphics[scale=0.50]{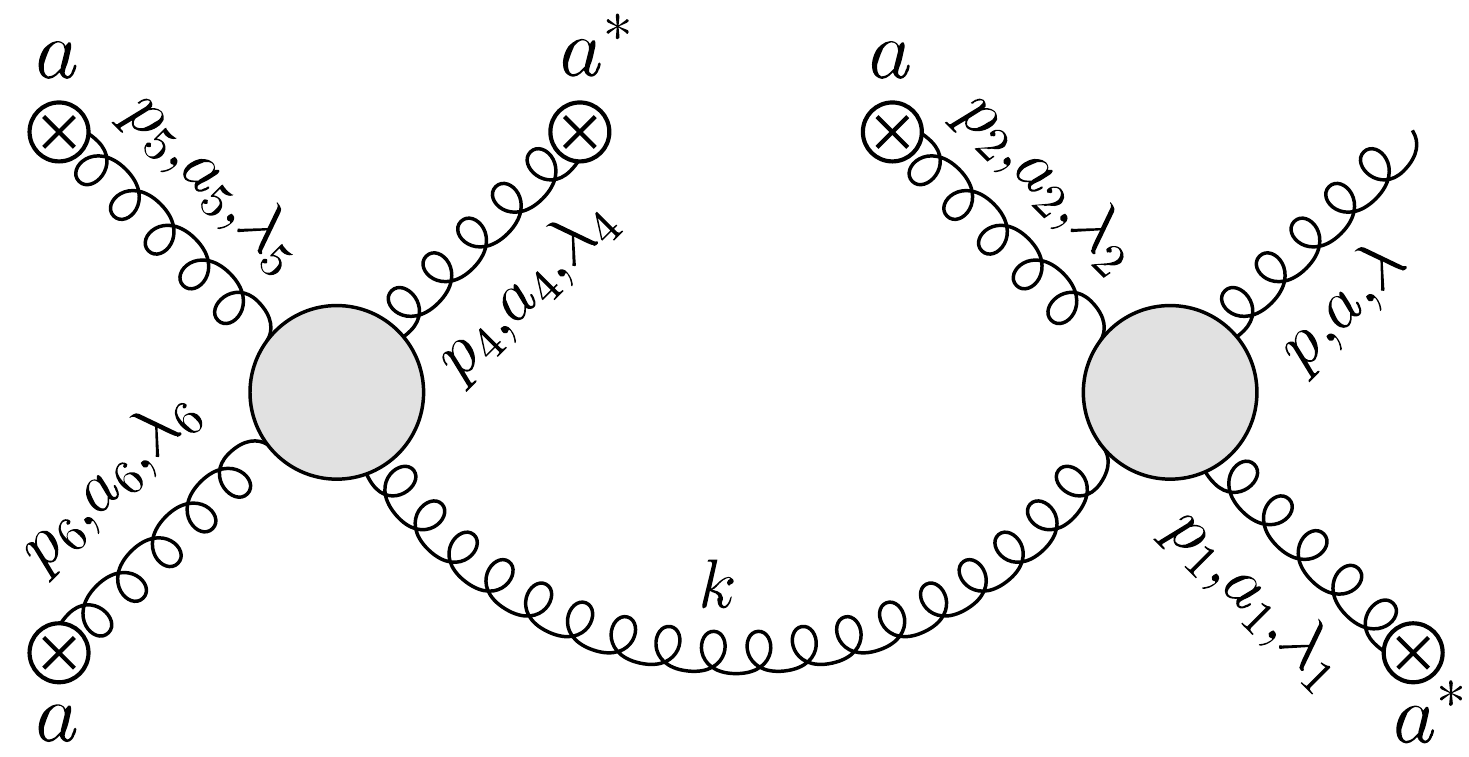}\\{\small (a)}
\end{center}
\end{minipage}
\begin{minipage}[b]{0.49\textwidth}
\begin{center}
\includegraphics[scale=0.5]{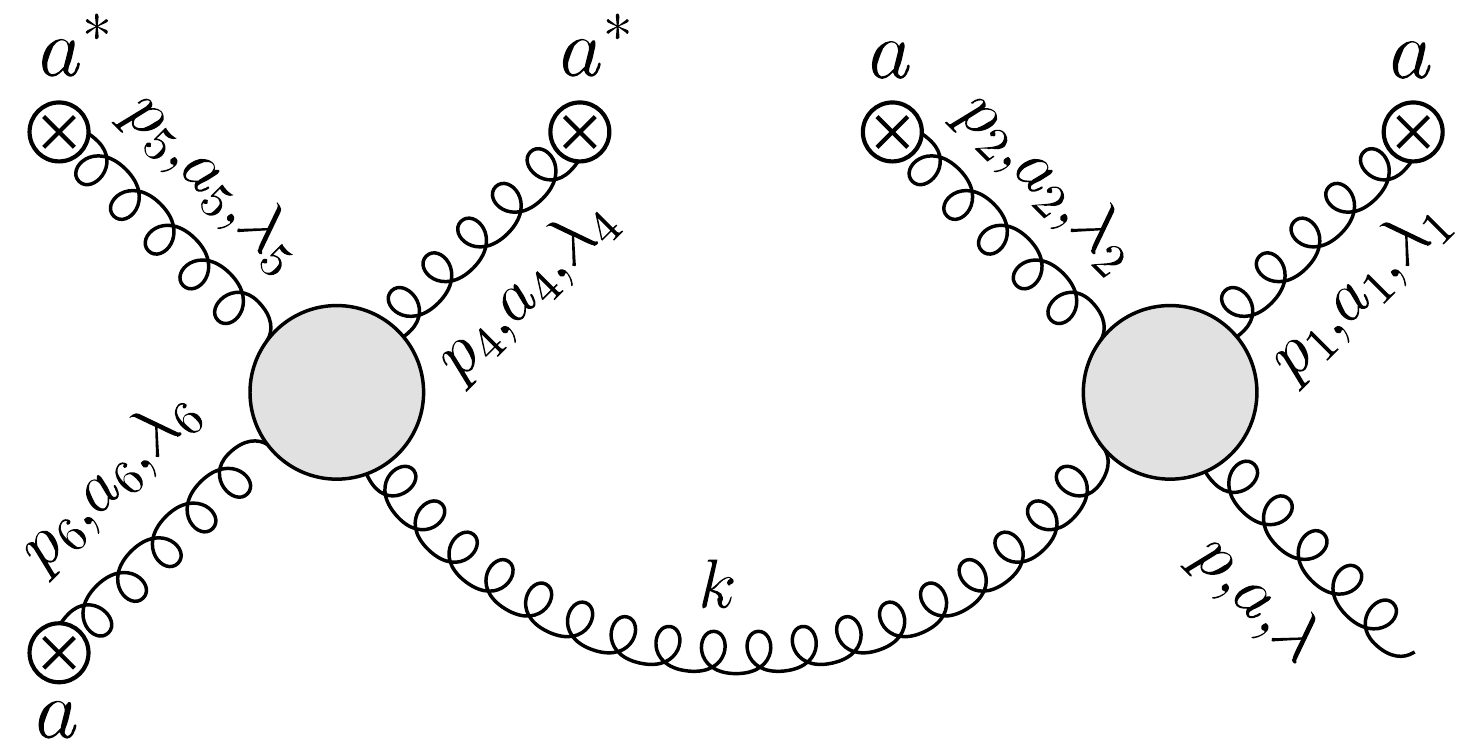}\\{\small (b)}
\end{center}
\end{minipage}
\caption{The contributions to $\delta a_{\bp}$ leading to (a) the loss terms and (b) the second gain term in the Boltzmann equation in Yang-Mills theory. The grey blob stands for the total amplitude for $2\to2$ scattering stripped off its polarization vectors (denoted by $\widetilde{\mcal{M}}$ in the text). In (a) the momentum $k$ is flowing to the right and the propagator is retarded while in (b) the momentum is flowing to the left and the propagator is advanced.}
\label{fig:dapcrossym}
\end{figure}

In order to derive the loss terms and the second gain term from the crossed contribution in \eqn{fpdotym} we need to calculate the $g^4$ contribution to $\delta a_{\bp}^{\lambda a}$. Regarding the loss term, and given the discussion below \eqn{sasf3f4} at the end of Sect.~\ref{sec:scalar}, it is not hard to understand that the sum of all possible contributing diagrams is given by
 \begin{align}
 \label{daplossym}
 \delta a_{\bp}^{\lambda a} = -
 \frac{1}{2 h_{\bp}}
 \int & \prod_{i} \frac{\dif^3 \bp_{i}}{h_{\bp_i}}\,
 (2\pi)^3 \delta^{(3)}(\overline{\Delta\bp})\,
 \frac{\rme^{\rmi (\overline{\Delta E} - \rmi \epsilon) x^0}}
 {\rmi(\overline{\Delta E} - \rmi \epsilon)}\,
 a^{\lambda_1 a_1*}_{\bp_1} a^{\lambda_2 a_2}_{\bp_2} 
 a^{\lambda_4 a_4*}_{\bp_4} a^{\lambda_5 a_5}_{\bp_5} 
 a^{\lambda_6 a_6}_{\bp_6}
 \nn & 
 \varepsilon^{\lambda*}_{\mu}(\bp)
 \varepsilon^{\lambda_1*}_{\mu_1}(\bp_1)
 \varepsilon^{\lambda_2}_{\mu_2}(\bp_2)
 \varepsilon^{\lambda_4*}_{\mu_4}(\bp_4)
 \varepsilon^{\lambda_5}_{\mu_5}(\bp_5)
 \varepsilon^{\lambda_6}_{\mu_6}(\bp_6)
 \nn*[0.2cm] &
 \widetilde{\mcal{M}}_{\nu\mu_4\mu_5\mu_6}^{b a_4a_5 a_6}
 (k p_4; p_5 p_6)\,
 G_{\rm R}^{\nu\rho}(k)\,
 \widetilde{\mcal{M}}_{\rho\mu_2\mu\mu_1}^{b a_2 a a_1 *}
 (k p_2;p p_1),
 \end{align}
with $k=p_5 + p_6 - p_4$ and where $\widetilde{\mcal{M}}$ is the total amplitude $\mcal{M}$ given by the sum of Eqs.~\eqref{m3g} and \eqref{m4g}, but stripped off its polarization vectors. Because $\widetilde{\mcal{M}}$ is imaginary, in the last factor we have let $\widetilde{\mcal{M}} \to -\widetilde{\mcal{M}}^*$ and this is the origin of the minus sign in \eqn{daplossym}. In Fig.~\ref{fig:dapcrossym}.a we show the diagrammatic representation of \eqn{daplossym}.

The steps to be followed now are of course almost identical to those in the scalar theory. A notable difference is the Lorentz structure of the propagator in \eqn{grprop}, which, since the propagator is eventually put on-shell, can be expressed as a sum over polarization vectors. More precisely one can write 
 \beq
 \label{proppol}
 -g^{\nu\rho} + 
 \frac{n^\nu k^{\rho} + n^{\rho}k^{\nu}}{n\cdot k} =
 \varepsilon_{\lambda''}^{\nu*}(\bk)\,
 \varepsilon_{\lambda''}^{\rho}(\bk),
 \eeq
where, as always, a summation over $\lambda''$ is understood. In total we now have eight polarization vectors in \eqn{daplossym} and they exactly combine with the $\widetilde{\mcal{M}}$'s there to give the corresponding standard amplitudes $\mcal{M}$. Then by closely following the calculation of the scalar theory, employing the ensemble average
\begin{align}
\label{factymcross}
\hspace*{-1.13cm}\big\langle a^{\lambda_1 a_1*}_{\bp_1} a^{\lambda_2 a_2}_{\bp_2} 
a^{\lambda_4 a_4*}_{\bp_4} 
a^{\lambda_5 a_5}_{\bp_5} 
a^{\lambda_6 a_6}_{\bp_6} 
a^{\lambda' a'*}_{\bp'}
\big\rangle
\!\to\! 
2\delta^{\lambda_1\lambda_5}
\delta^{\lambda_2\lambda_4}
\delta^{\lambda_6\lambda'}
\delta^{a_1 a_5}
\delta^{a_2 a_4}
\delta^{a_6 a'}
\delta^{(3)}_{\bp_1\bp_5}
\delta^{(3)}_{\bp_2\bp_4} 
\delta^{(3)}_{\bp_6\bp'}
f_{\bp'}f_{\bp_1}f_{\bp_2} 
\end{align}
and using \eqn{mmstar} we arrive at the loss term
 \beq\label{beymb} 
 \dot f_{\bm p}\big|^B =  
 -\frac{1}{4 E_{\bp}} 
 \int \widetilde{\dif \bp_1}\,
 \widetilde{\dif \bp_2}\,
 \widetilde{\dif \bp_3}\,
 (2\pi)^4 \delta^{(4)}(\Delta p)\,
 \frac{|\mcal{M}|^2_{\rm YM}}{2(N_c^2-1)}\, 
 \big[ f_{\bp}f_{\bp_1}f_{\bp_2} +
 f_{\bp}f_{\bp_1}f_{\bp_3} \big].
 \eeq
It is more than obvious that if we choose the combination $a_1 a_2 a_4^* a_5^* a_6$, instead of $a_1^* a_2 a_4^* a_5 a_6$, we shall arrive at the second gain term and the respective diagram is shown in Fig.~\ref{fig:dapcrossym}.b. Putting everything together we find the Boltzmann equation in classical Yang-Mills theory, that is
 \beq
 \label{beym} 
 \dot f_{\bm p}=  
 \frac{1}{4 E_{\bp}} 
 \int \widetilde{\dif \bp_1}\,
 \widetilde{\dif \bp_2}\,
 \widetilde{\dif \bp_3}\,
 (2\pi)^4 \delta^{(4)}(\Delta p)\,
 \frac{|\mcal{M}|^2_{\rm YM}}{2(N_c^2-1)}\, 
 \big[ f_{\bp_2}f_{\bp_3} \big(f_{\bp_1} + f_{\bp}\big)   
 -f_{\bp}f_{\bp_1}  \big(f_{\bp_2} + f_{\bp_3}\big) \big],
 \eeq
with $|\mcal{M}|^2_{\rm YM}$ given earlier in \eqn{m2ym}. Notice that the combination which appears in the integrand is really the amplitude squared averaged over the color and polarization of the measured gluon and summed over the colors and polarizations of the remaining three gluons. As in the scalar field theory, a factor 1/2 in front of the integral is a symmetry factor due to the fact that particles 2 and 3 are identical.

Before closing let us repeat here an observation made in \cite{Mueller:2002gd}. At the level of the classical approximation, since $f_{\bp} \gg 1$, we can assume a modified definition of the occupation number by replacing $f_{\bp}$ in the r.h.s.~of \eqn{astaraym} with $f_{\bp} + 1/2$. Then, such a replacement is carried to all occupations numbers appearing in the collision integral of the Boltzmann equation and one sees that the cubic in $f$ terms remain unaltered as they should. Interestingly enough, the generated quadratic in $f$ terms are exactly those present in the more general Boltzmann equation which is valid for all values of $f_{\bp}$ and is given in \eqn{beintro}. However, such a replacement also gives rise to linear in $f$ terms which are absent from \eqn{beintro}.

\begin{acknowledgments}
D.N.T.~would like to thank Edmond Iancu for discussions and suggestions. V.M.~is supported by the Indiana University Collaborative Research Grant (IUCRG). The work of A.H.M.~is supported, in part, by the US Department of Energy. All figures were created with Jaxodraw \cite{Binosi:2003yf}.
\end{acknowledgments}

%\appendix*
%\section{\label{sec:app}To be written}

%\bibliographystyle{utcaps}
%\bibliography{../../References/refs}

\begin{thebibliography}{10}

\bibitem{Baym:1990uj}
G.~Baym, H.~Monien, C.~Pethick, and D.~Ravenhall, ``{Transverse Interactions
  and Transport in Relativistic Quark - Gluon and Electromagnetic Plasmas},''
\href{http://dx.doi.org/10.1103/PhysRevLett.64.1867}{{\em Phys.Rev.Lett.}
  {\bfseries 64} (1990) 1867--1870}.
%%CITATION = PRLTA,64,1867;%%.

\bibitem{Jeon:1995zm}
S.~Jeon and L.~G. Yaffe, ``{From quantum field theory to hydrodynamics:
  Transport coefficients and effective kinetic theory},''
  \href{http://dx.doi.org/10.1103/PhysRevD.53.5799}{{\em Phys.Rev.} {\bfseries
  D53} (1996) 5799--5809},
\href{http://arxiv.org/abs/hep-ph/9512263}{{\ttfamily arXiv:hep-ph/9512263
  [hep-ph]}}.
%%CITATION = HEP-PH/9512263;%%.

\bibitem{Arnold:2000dr}
P.~B. Arnold, G.~D. Moore, and L.~G. Yaffe, ``{Transport coefficients in high
  temperature gauge theories. 1. Leading log results},''
  \href{http://dx.doi.org/10.1088/1126-6708/2000/11/001}{{\em JHEP} {\bfseries
  0011} (2000) 001},
\href{http://arxiv.org/abs/hep-ph/0010177}{{\ttfamily arXiv:hep-ph/0010177
  [hep-ph]}}.
%%CITATION = HEP-PH/0010177;%%.

\bibitem{Arnold:2003zc}
P.~B. Arnold, G.~D. Moore, and L.~G. Yaffe, ``{Transport coefficients in high
  temperature gauge theories. 2. Beyond leading log},''
  \href{http://dx.doi.org/10.1088/1126-6708/2003/05/051}{{\em JHEP} {\bfseries
  0305} (2003) 051},
\href{http://arxiv.org/abs/hep-ph/0302165}{{\ttfamily arXiv:hep-ph/0302165
  [hep-ph]}}.
%%CITATION = HEP-PH/0302165;%%.

\bibitem{Selikhov:1993ns}
A.~Selikhov and M.~Gyulassy, ``{Color diffusion and conductivity in a quark -
  gluon plasma},'' \href{http://dx.doi.org/10.1016/0370-2693(93)90341-E}{{\em
  Phys.Lett.} {\bfseries B316} (1993) 373--380},
\href{http://arxiv.org/abs/nucl-th/9307007}{{\ttfamily arXiv:nucl-th/9307007
  [nucl-th]}}.
%%CITATION = NUCL-TH/9307007;%%.

\bibitem{Heiselberg:1994px}
H.~Heiselberg, ``{Color, spin and flavor diffusion in quark - gluon plasmas},''
  \href{http://dx.doi.org/10.1103/PhysRevLett.72.3013}{{\em Phys.Rev.Lett.}
  {\bfseries 72} (1994) 3013--3016},
\href{http://arxiv.org/abs/hep-ph/9401317}{{\ttfamily arXiv:hep-ph/9401317
  [hep-ph]}}.
%%CITATION = HEP-PH/9401317;%%.

\bibitem{Bodeker:1998hm}
D.~Bodeker, ``{On the effective dynamics of soft nonAbelian gauge fields at
  finite temperature},''
  \href{http://dx.doi.org/10.1016/S0370-2693(98)00279-2}{{\em Phys.Lett.}
  {\bfseries B426} (1998) 351--360},
\href{http://arxiv.org/abs/hep-ph/9801430}{{\ttfamily arXiv:hep-ph/9801430
  [hep-ph]}}.
%%CITATION = HEP-PH/9801430;%%.

\bibitem{Arnold:1998cy}
P.~B. Arnold, D.~T. Son, and L.~G. Yaffe, ``{Effective dynamics of hot, soft
  nonAbelian gauge fields. Color conductivity and log(1/alpha) effects},''
  \href{http://dx.doi.org/10.1103/PhysRevD.59.105020}{{\em Phys.Rev.}
  {\bfseries D59} (1999) 105020},
\href{http://arxiv.org/abs/hep-ph/9810216}{{\ttfamily arXiv:hep-ph/9810216
  [hep-ph]}}.
%%CITATION = HEP-PH/9810216;%%.

\bibitem{Blaizot:1999xk}
J.-P. Blaizot and E.~Iancu, ``{A Boltzmann equation for the QCD plasma},''
  \href{http://dx.doi.org/10.1016/S0550-3213(99)00341-7}{{\em Nucl.Phys.}
  {\bfseries B557} (1999) 183--236},
\href{http://arxiv.org/abs/hep-ph/9903389}{{\ttfamily arXiv:hep-ph/9903389
  [hep-ph]}}.
%%CITATION = HEP-PH/9903389;%%.

\bibitem{Blaizot:2001nr}
J.-P. Blaizot and E.~Iancu, ``{The Quark gluon plasma: Collective dynamics and
  hard thermal loops},''
  \href{http://dx.doi.org/10.1016/S0370-1573(01)00061-8}{{\em Phys.Rept.}
  {\bfseries 359} (2002) 355--528},
\href{http://arxiv.org/abs/hep-ph/0101103}{{\ttfamily arXiv:hep-ph/0101103
  [hep-ph]}}.
%%CITATION = HEP-PH/0101103;%%.

\bibitem{Kadanoff:1962aaaa}
L.~Kadanoff and G.~Baym, {\em {Quantum statistical mechanics}}.
\newblock {Benjamin}, 1962.

\bibitem{Calzetta:1986cq}
E.~Calzetta and B.~Hu, ``{Nonequilibrium Quantum Fields: Closed Time Path
  Effective Action, Wigner Function and Boltzmann Equation},''
\href{http://dx.doi.org/10.1103/PhysRevD.37.2878}{{\em Phys.Rev.} {\bfseries
  D37} (1988) 2878}.
%%CITATION = PHRVA,D37,2878;%%.

\bibitem{Mrowczynski:1989bu}
S.~Mrowczynski and P.~Danielewicz, ``{Green Function Approach to Transport
  Theory of Scalar Fields},''
\href{http://dx.doi.org/10.1016/0550-3213(90)90194-I}{{\em Nucl.Phys.}
  {\bfseries B342} (1990) 345--380}.
%%CITATION = NUPHA,B342,345;%%.

\bibitem{Mrowczynski:1992hq}
S.~Mrowczynski and U.~W. Heinz, ``{Towards a relativistic transport theory of
  nuclear matter},''
\href{http://dx.doi.org/10.1006/aphy.1994.1001}{{\em Annals Phys.} {\bfseries
  229} (1994) 1--54}.
%%CITATION = APNYA,229,1;%%.

\bibitem{Jeon:1994if}
S.~Jeon, ``{Hydrodynamic transport coefficients in relativistic scalar field
  theory},'' \href{http://dx.doi.org/10.1103/PhysRevD.52.3591}{{\em Phys.Rev.}
  {\bfseries D52} (1995) 3591--3642},
\href{http://arxiv.org/abs/hep-ph/9409250}{{\ttfamily arXiv:hep-ph/9409250
  [hep-ph]}}.
%%CITATION = HEP-PH/9409250;%%.

\bibitem{Blaizot:1993be}
J.~P. Blaizot and E.~Iancu, ``{Soft collective excitations in hot gauge
  theories},'' \href{http://dx.doi.org/10.1016/0550-3213(94)90486-3}{{\em
  Nucl.Phys.} {\bfseries B417} (1994) 608--673},
\href{http://arxiv.org/abs/hep-ph/9306294}{{\ttfamily arXiv:hep-ph/9306294
  [hep-ph]}}.
%%CITATION = HEP-PH/9306294;%%.

\bibitem{Iancu:1998sg}
E.~Iancu, ``{Effective theory for real time dynamics in hot gauge theories},''
\href{http://dx.doi.org/10.1016/S0370-2693(98)00772-2}{{\em Phys.Lett.}
  {\bfseries B435} (1998) 152--158}.
%%CITATION = PHLTA,B435,152;%%.

\bibitem{Calzetta:2008aaaa}
E.~A. Calzetta and B.-L. Hu, {\em {Nonequilibrium quantum field theory}}.
\newblock {Cambridge University Press}, 2008.

\bibitem{Litim:1999ns}
D.~F. Litim and C.~Manuel, ``{Mean field dynamics in nonAbelian plasmas from
  classical transport theory},''
  \href{http://dx.doi.org/10.1103/PhysRevLett.82.4981}{{\em Phys.Rev.Lett.}
  {\bfseries 82} (1999) 4981--4984},
\href{http://arxiv.org/abs/hep-ph/9902430}{{\ttfamily arXiv:hep-ph/9902430
  [hep-ph]}}.
%%CITATION = HEP-PH/9902430;%%.

\bibitem{Litim:2001db}
D.~F. Litim and C.~Manuel, ``{Semiclassical transport theory for nonAbelian
  plasmas},'' \href{http://dx.doi.org/10.1016/S0370-1573(02)00015-7}{{\em
  Phys.Rept.} {\bfseries 364} (2002) 451--539},
\href{http://arxiv.org/abs/hep-ph/0110104}{{\ttfamily arXiv:hep-ph/0110104
  [hep-ph]}}.
%%CITATION = HEP-PH/0110104;%%.

\bibitem{Mueller:2002gd}
A.~Mueller and D.~Son, ``{On the Equivalence between the Boltzmann equation and
  classical field theory at large occupation numbers},''
  \href{http://dx.doi.org/10.1016/j.physletb.2003.12.047}{{\em Phys.Lett.}
  {\bfseries B582} (2004) 279--287},
\href{http://arxiv.org/abs/hep-ph/0212198}{{\ttfamily arXiv:hep-ph/0212198
  [hep-ph]}}.
%%CITATION = HEP-PH/0212198;%%.

\bibitem{Jeon:2004dh}
S.~Jeon, ``{The Boltzmann equation in classical and quantum field theory},''
  \href{http://dx.doi.org/10.1103/PhysRevC.72.014907}{{\em Phys.Rev.}
  {\bfseries C72} (2005) 014907},
\href{http://arxiv.org/abs/hep-ph/0412121}{{\ttfamily arXiv:hep-ph/0412121
  [hep-ph]}}.
%%CITATION = HEP-PH/0412121;%%.

\bibitem{Blaizot:2011xf}
J.-P. Blaizot, F.~Gelis, J.-F. Liao, L.~McLerran, and R.~Venugopalan,
  ``{Bose--Einstein Condensation and Thermalization of the Quark Gluon
  Plasma},'' \href{http://dx.doi.org/10.1016/j.nuclphysa.2011.10.005}{{\em
  Nucl.Phys.} {\bfseries A873} (2012) 68--80},
\href{http://arxiv.org/abs/1107.5296}{{\ttfamily arXiv:1107.5296 [hep-ph]}}.
%%CITATION = ARXIV:1107.5296;%%.

\bibitem{Epelbaum:2011pc}
T.~Epelbaum and F.~Gelis, ``{Role of quantum fluctuations in a system with
  strong fields: Spectral properties and Thermalization},''
  \href{http://dx.doi.org/10.1016/j.nuclphysa.2011.09.019}{{\em Nucl.Phys.}
  {\bfseries A872} (2011) 210--244},
\href{http://arxiv.org/abs/1107.0668}{{\ttfamily arXiv:1107.0668 [hep-ph]}}.
%%CITATION = ARXIV:1107.0668;%%.

\bibitem{Epelbaum:2014yja}
T.~Epelbaum, F.~Gelis, and B.~Wu, ``{Non-renormalizability of the classical
  statistical approximation},''
\href{http://arxiv.org/abs/1402.0115}{{\ttfamily arXiv:1402.0115 [hep-ph]}}.
%%CITATION = ARXIV:1402.0115;%%.

\bibitem{Berges:2013fga}
J.~Berges, K.~Boguslavski, S.~Schlichting, and R.~Venugopalan, ``{Universal
  attractor in a highly occupied non-Abelian plasma},''
\href{http://arxiv.org/abs/1311.3005}{{\ttfamily arXiv:1311.3005 [hep-ph]}}.
%%CITATION = ARXIV:1311.3005;%%.

\bibitem{Berges:2013lsa}
J.~Berges, K.~Boguslavski, S.~Schlichting, and R.~Venugopalan, ``{Basin of
  attraction for turbulent thermalization and the range of validity of
  classical-statistical simulations},''
\href{http://arxiv.org/abs/1312.5216}{{\ttfamily arXiv:1312.5216 [hep-ph]}}.
%%CITATION = ARXIV:1312.5216;%%.

\bibitem{Baier:2000sb}
R.~Baier, A.~H. Mueller, D.~Schiff, and D.~Son, ``{'Bottom up' thermalization
  in heavy ion collisions},''
  \href{http://dx.doi.org/10.1016/S0370-2693(01)00191-5}{{\em Phys.Lett.}
  {\bfseries B502} (2001) 51--58},
\href{http://arxiv.org/abs/hep-ph/0009237}{{\ttfamily arXiv:hep-ph/0009237
  [hep-ph]}}.
%%CITATION = HEP-PH/0009237;%%.

\bibitem{Gelis:2013rba}
T.~Epelbaum and F.~Gelis, ``{Pressure isotropization in high energy heavy ion
  collisions},'' \href{http://dx.doi.org/10.1103/PhysRevLett.111.232301}{{\em
  Phys.Rev.Lett.} {\bfseries 111} (2013) 232301},
\href{http://arxiv.org/abs/1307.2214}{{\ttfamily arXiv:1307.2214 [hep-ph]}}.
%%CITATION = ARXIV:1307.2214;%%.

\bibitem{Srednicki:2007qs}
M.~Srednicki, {\em {Quantum field theory}}.
\newblock {Cambridge University Press},
2007.
\newblock
%%CITATION = INSPIRE-752478;%%.

\bibitem{Binosi:2003yf}
D.~Binosi and L.~Theussl, ``{JaxoDraw: A Graphical user interface for drawing
  Feynman diagrams},'' \href{http://dx.doi.org/10.1016/j.cpc.2004.05.001}{{\em
  Comput.Phys.Commun.} {\bfseries 161} (2004) 76--86},
\href{http://arxiv.org/abs/hep-ph/0309015}{{\ttfamily arXiv:hep-ph/0309015
  [hep-ph]}}.
%%CITATION = HEP-PH/0309015;%%.

\end{thebibliography}

\providecommand{\href}[2]{#2}\begingroup\raggedright\endgroup

\end{document}